\documentclass{report}

\usepackage[breaklinks=true]{hyperref}
\usepackage{breakcites}

\usepackage{array}  

\usepackage{blindtext}

\usepackage{enumerate}
\usepackage[shortlabels]{enumitem}

\usepackage{graphicx}

\usepackage{amsfonts}

\usepackage{bm}

\usepackage{mathtools}

\usepackage{ragged2e}

\usepackage[export]{adjustbox}

\usepackage{ amssymb }

 \usepackage[utf8]{inputenc}
 \usepackage[T1]{fontenc}

\usepackage[section]{placeins}

\usepackage[numbers,sort&compress]{natbib}

 \usepackage{lmodern}

\title{\Large \textbf{\fontfamily{qcr}\selectfont{Machine Learning and Deep Learning Techniques used in Cybersecurity and Digital Forensics : a Review}}}
\author{Jaouhar Fattahi, Ph.D}
\date{\today}
\begin{document}

\newcolumntype{C}[1]{>{\centering\let\newline\\\arraybackslash\hspace{0pt}}m{#1}}
\newcolumntype{L}[1]{>{\let\newline\\\arraybackslash\hspace{0pt}}m{#1}}
\def\Z{\vphantom{\parbox[c]{1cm}{\Huge Something Long}}}
\def\N{\vphantom{\parbox[c]{1cm}{}}}
\bigskip
\def\ZZ{\vphantom{\parbox[c]{1cm}{\Large Something Long}}}

\maketitle

\begin{abstract}
In the paced realms of cybersecurity and digital forensics machine learning (ML) and deep learning (DL) have emerged as game changing technologies that introduce methods to identify stop and analyze cyber risks. This review presents an overview of the ML and DL approaches used in these fields showcasing their advantages drawbacks and possibilities. It covers a range of AI techniques used in spotting intrusions in systems and classifying malware to prevent cybersecurity attacks, detect anomalies and enhance resilience. This study concludes by highlighting areas where further research is needed and suggesting ways to create transparent and scalable ML and DL solutions that are suited to the evolving landscape of cybersecurity and digital forensics. 
\end{abstract}

\renewcommand\thesection{\arabic{section}}
\newpage
\tableofcontents\clearpage
\listoffigures\clearpage
\listoftables

\newpage

\newpage
\section{Introduction}

In this age of technology and digital advancements cybersecurity has become a priority due to the fast paced growth of technology and the growing dependence on digital platforms. This reliance has opened up avenues for advancement and creativity. It has also brought about risks and dangers that pose a challenge to safeguarding digital resources and confidential data. Digital forensics works hand in hand with cybersecurity by looking into and solving incidents that happen on; it concentrates on collecting and safeguard proof and then examining it further to find answers. This process is vital in identifying cybercrimes like identity theft and fraud while also making sure that those responsible are held accountable in the world. The introduction of machine learning (ML) and deep learning (DL) techniques into investigations have expanded its functionalities by automating the analysis of data sets efficiently. This helps in better spotting harmful software and getting deeper insights, into intricate cyber events. In this study we are going to review common machine learning and deep learning techniques used in cybersecurity and digital forensics applications.

\newpage

 \section{Standard Machine Learning techniques and algorithms}     

\subsection{Machine Learning}

Machine Learning is an area of artificial intelligence that involves the design, analysis, development, and implementation of methods that allow a machine to evolve through a systematic process, which allows it to accomplish tasks that are difficult to perform by conventional algorithmic means without using explicit instructions. This process roughly consists of the following seven sequential steps.

\begin{enumerate}

\item  Collecting data: this step is extremely important because the quality and quantity of data the data scientist collects will directly affect the quality of the predictive model. A good data collection must ensure that data belonging to one class or another is balanced (i.e. reflecting the reality). The collection should also be as comprehensive as possible and should not focus only on the general case.  It must also provide a sufficient number of samples that can provide reliable results. The more samples there are, the more generalizable the predictive model is.

\item Investigating data: this step aims at understanding the characteristics of the collected data and probably pre-processing them prior to being used to build the model. In this step, it is a good practice to perform some checks on data by printing a data summary containing the number of samples, the number of classes, the number of samples pertaining to a class, the data distribution, etc. Using visualization tools is also recommended to pick up relevant relationships between different variables and to discover if the data are balanced or not and if there are missing data or not. The data scientist needs also to perform some relevant and useful  measurements like calculating the correlation between features and the output to determine which features are important and which ones are less important, as well as whether the features are linearly correlated or not. 

\item Preparing data: a data scientist needs to split the data in at least two parts. The first part will serve to train the model. The second part will serve to evaluate the model performance. In a wide range of cases, data needs other types of treatment like normalization, error correction,  tokenization, vectorization, data type conversion, data de-duping, etc.

\item Choosing the model: the choice of the best model for a given problem depends on the size, quality, and nature of the data. Even the most experienced data scientists cannot always know which algorithm will work best before they try it. However, there are some models that are better suited to certain types of problems. For instance, some are very well suited to image data, others to sequences (such as text or music), some to digital data, others to particular textual data. 

\item Training the model: To build a model, one first tries to find the optimal model using the training data. This consists in using a metric called a loss function that uses numerical optimization techniques. The loss function determines to what extent the modelling of the problem, which is an approximation of reality, loses information in relation to the reality observed through the data samples. Structured, clean, and consistent data samples are worth gold. It is also important to have the right amount of data. To ensure  better learning dynamics while making reasonable use of available resources, it is common to split the data into packets, called batches. When we have gone through all the batches, we say that we have run an epoch.

\item Evaluating the model:  model evaluation (or validation or testing) consists in running it with data that have never been used for training. This allows to observe how the model could work with data that it has not yet explored. Many metrics are widely used to determine how well the built model behaves, such as accuracy, precision, recall and F1 Score. The confusion matrix is also widely used to determine false positives, false negatives, true positives and true negatives. To provide a visual touch, Receiver Operating Characteristic (ROC) curves could also be used. 

\item Tuning the model parameters: once the model evaluation is completed, it is possible that its performance may be unsatisfactory and that it could improve further. This can be done by tweaking its parameters. These parameters can be related to initialization, regularization, optimization, learning rate, batches sizes, number of epochs, or even to the model architecture.

\end{enumerate}

\subsubsection{Supervised learning}     

Learning is said to be supervised when the data entering the process are already categorized (labeled) and algorithms must use them to predict an outcome so that they can proceed the same way when the data are no longer categorized. In other words, in supervised learning one has prior knowledge of what the output values of the samples should be. 

\subsubsection{Unsupervised learning}      

Unsupervised learning is much more complex since the system has to detect similarities in the data it receives and organize them accordingly. Unsupervised learning therefore has no labeled outcomes. Its objective is thus to deduce the natural structure present in a set of data points.  In other words, in unsupervised learning one does not have prior knowledge of what the output values of the samples might be. 

\subsubsection{Classification}      

Classification is a family of supervised methods used to categorize data using an algorithm. It requires that the data points have discrete values. The learning data are grouped in categories (classes). Then, the new input data are classified according to the learned data. Classification produces discrete values.

\subsubsection{Regression}      

Regression is a supervised method for understanding the relationships between a variable of interest and explanatory variables. Regression usually requires that the data points have continuous values. In a regression, first, the factors (i.e. independent variables) are found. Then, coefficients (i.e. multipliers) with independent variables are calculated to minimize differences between true and predicted values. Finally, a formula for the regression is set. This formula is used to predict the dependent variable (i.e. one wants to measure) from independent variables (i.e. one thinks the target measure hinges on). 

\subsubsection{Clustering}      

Clustering is an unsupervised technique that consists in grouping the data into homogeneous groups called clusters, so that the elements within the same cluster are similar, and the elements belonging to two different clusters are not similar.

\subsubsection{Reinforcement learning}      

Reinforcement Learning is a method that consists in letting the algorithm learn from its own errors. The algorithm begins by making totally random decisions and then develops its own method to correctly accomplish the task entrusted to it. That is, an agent in a current state $S$ learns from its environment by interacting with it through actions. Once an action $A$ is performed, the environment returns a new state $S'$ along with an associated reward $R$, which can be positive (for a good action) or negative (for a bad action). This process is repeated until the agent decision reaches a confirmed maturity translated into a long and steady behaviour of good actions. Thus, this learning method has the advantage of allowing the machine to be more creative.




\subsubsection{Common problems}        

\begin{itemize}

\item Underfitting: this problem happens when the model cannot capture enough patterns from data. This may be the case when one uses a model for a problem of an incompatible nature (e.g. a linear regression for a non-linear problem, a simple Artificial Neural Network  (ANN) for a sequential problem, an Recurrent Neural Network (RNN) without memory for problems needing past information, etc.). When underfitting is observed, the model accuracy is generally low.

\item Overfitting: this is the opposite of the under-fitting problem. It happens when the learning clings too closely, or even exactly, to a particular set of data. We say that the model learns the existing data \textit{by heart}. A model in an underfitting (or overlearning) situation has difficulty generalizing the characteristics of the data and loses its predictive capabilities on new samples.

\begin{figure}[h]

\centering
 \includegraphics[scale=0.3, frame]{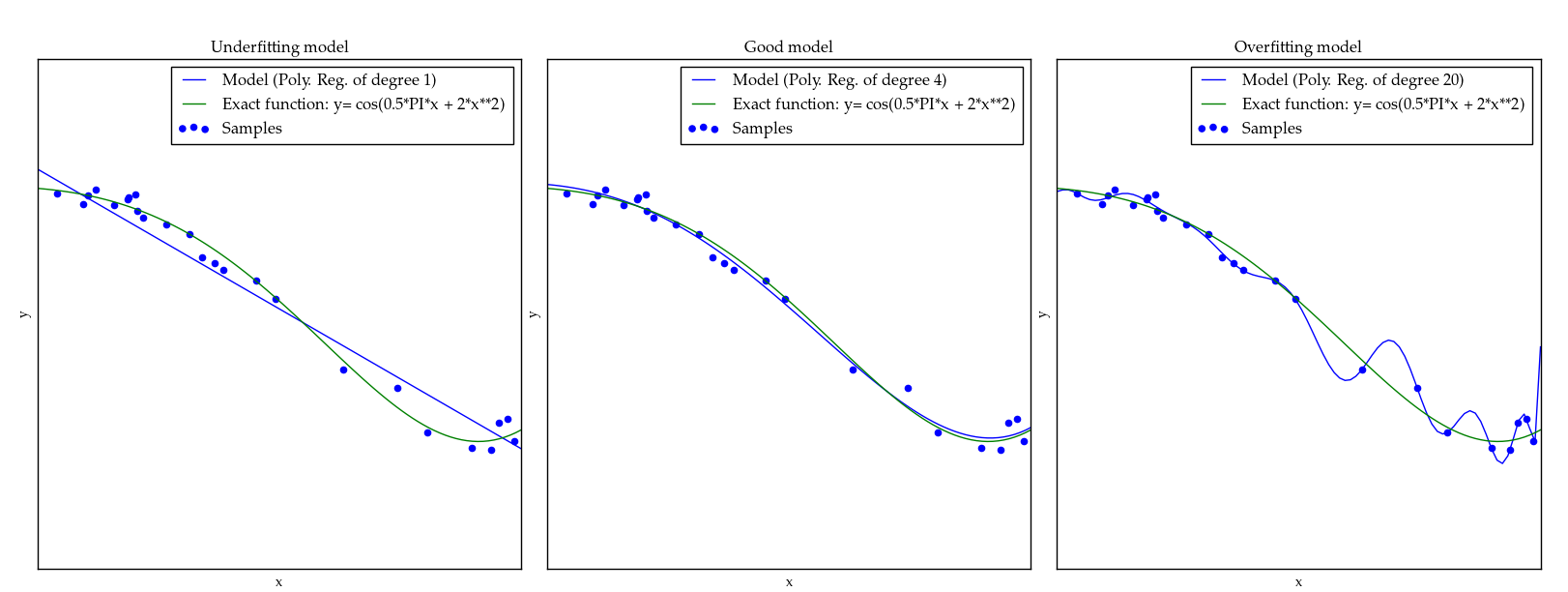}

\caption{Model underfitting and overfitting}
\label{fig:fitProb}
\end{figure}  

\item Missing data: real-world data are not always neat and consistent. Generally, they are prone to being incomplete, noisy, and incongruous, and it is important for the analyst to handle them by either populating in the missing values or removing them as they can induce prediction or classification mistakes. There are many possible reasons why they happen, ranging from human mistakes in data acquisition to imprecise sensor measurements by way of software glitches in the data handling process. A common way to correct this is to replace missing values with the mean, median, or mode values. They can also be attributed the value given by a regression model, such as linear regression, or the value having the majority of votes using a $k$-Nearest Neighbours ($k$-NN) model.

\item  Vanishing gradient: the problem of the vanishing gradient arises when one trains a neural network model using gradient optimization techniques. This is one of the major problems in Machine Learning and Deep Learning because it usually leads to a very long training time or even to a non-convergence towards the correct minimum. It is manifested by the fact that the gradients tend to become smaller and smaller as one keeps moving backward in the neural network during the back-propagation phase until the model ceases learning from data. This problem, which we will present in more detail in this document, is addressed in several ways, such as by using appropriate activation functions, or models with short memory such as Long Short-Term Memory (LSTM) models and Gated Recurrent Units (GRU) models.


\item Exploding gradient: gradient explosion occurs when large error gradients build up which results in huge updates to the weights of the model during the training phase. When gradient amplitudes accumulate, network instability generally results, which may lead to bad prediction outcomes. This problem is usually handled using clipping techniques to keep gradients stable.  With these techniques, a predefined gradient threshold is used, and then the gradients that are above this threshold are rescaled.


\item Initialization: initialization can have a direct effect on  model convergence. It can speed up training or it can generate awkward  problems. It may generate exploding or vanishing gradients. Some guide lines and techniques \cite{VishnuKakaraparthi,Kerasinitializers} are proposed to provide suitable initial values to a given model. This still remains an active research area though.

\end{itemize}




\subsection{Machine Learning Algorithms}        

\subsubsection{Linear Regressions}

Linear regression (LinR) was first used in statistics to understand the link between input and output of numerical variables. It is used in Machine Learning as a model to predict the output from the inputs under the assumption of a linear relationship between them. LinR is probably the simplest Machine Learning model.

For a single input and $x$ a single output $y$, the model consists in a line having the following equation:

$$y = b_0 + b_1 x$$

For multiple inputs,  the model consists in a hyperplane having the following equation

$$y = b_0 + b_1 x_1 + b_2 x_2 + ... + b_n x_n$$  

Various techniques can be used to prepare the linear regression equation from a training dataset  and hence to find the most adequate coefficients for the problem. The most widely used is the standard ordinary least (sum of) squares method. For a simple LinR, this method consists in minimizing the error observed between the predicted and the true outputs of the data set, {given by the following equation:


\begin{center}
\begin{tabular}{lll}

    $E(a,b)$ & $=$ &  $\displaystyle \sum_{i=1}^{n}  (y_i - \hat{y}_i)^2$\\
    & $=$ &  $\displaystyle \sum_{i=1}^{n}  (y_i - (\hat{b}_0 + \hat{b_1} x_i))^2$\\

\end{tabular}
\end{center}

where $\hat{y}_i$ is the actual output for the example $i$.

}

\begin{figure}[h]
\centering
\includegraphics[scale=0.5]{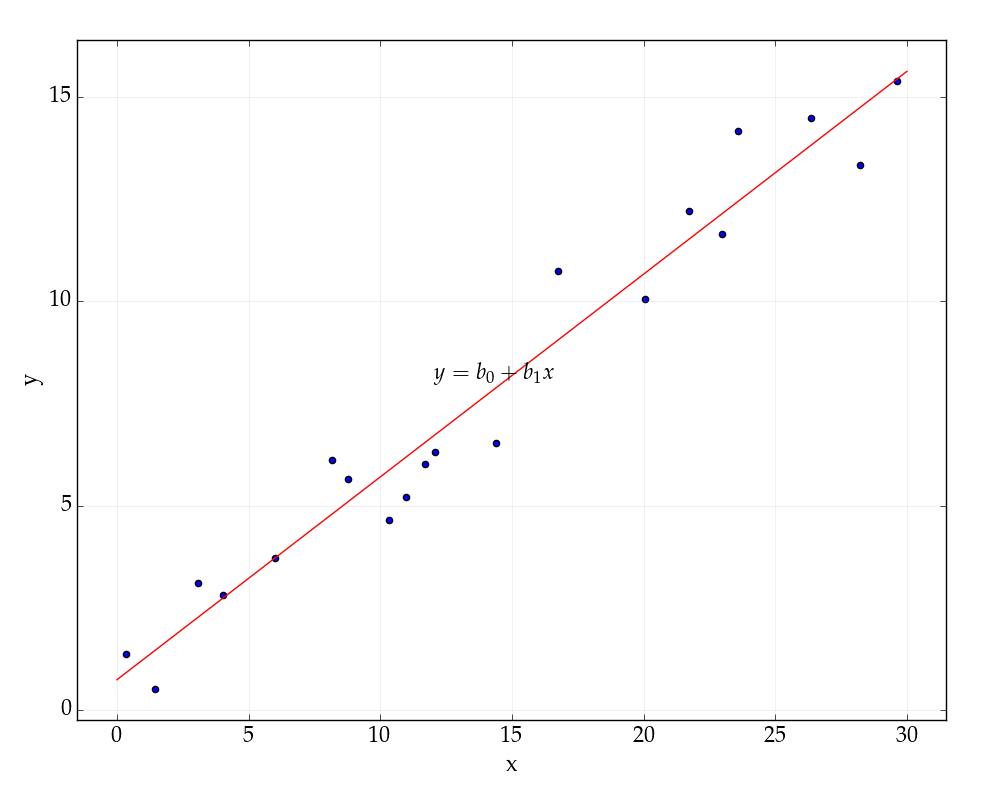}
\caption{Linear Regression}
\label{fig:LinRegression}
\end{figure}  

To determine the coefficients $\hat{b}_0$ and $\hat{b}_1$, the following steps are performed.

\begin{enumerate}
\item Compute the partial derivative of  $E(a,b)$ with respect to $\hat{b}_0$ and $\hat{b}_1$;
\item Set each partial derivative to zero;
\item Solve the resulting system of equations.
\end{enumerate}

This results in the following values:

$$
 \left\{
    \begin{array}{ll}
        \hat{b}_0 = & \bar{y} -\hat{b}_1\bar{x} \\
\\
\\
        \hat{b}_1 = & \displaystyle \frac{\displaystyle\sum_{i=1}^{n} (x_i - \bar{x})(y_i -\bar{y})}{\displaystyle\sum_{i=1}^{n} (x_i - \bar{x})^2} \\
    \end{array}
\right.
$$

where $\bar{x}$ is the mean value of the inputs and $\bar{y}$ is the mean value of the outputs. 

These two values are used as estimators for $b_0$ and $b_1$ respectively.


With multiple linear regression (i.e. linear regression with many predictors), $k$ independent variables and $k+1$ regression coefficient are involved. The steps to calculate these coefficients are similar to simple regression ones, but one must solve $k+1$ equations with $k+1$ unknowns to find the least-squares solution.  This involves a few matrix operations like matrix multiplications and matrix inversions.

It is worth mentioning that there is another type of regression called polynomial regression in which the regression function can involve terms such as $b_i x_j^k x_p^h$. This regression could be seen as a linear regression by performing a variable change and considering the term $x_j^k x_p^h$ as a new predictor.

To know how well a linear regression model fits the data, we calculate the coefficient of determination (also called $R$-squared) given by:

$$ R^2 = \displaystyle \frac{\displaystyle\sum_{i=1}^{n}(\hat{y}_i -\bar{y})^2}{\displaystyle\sum_{i=1}^{n} (y_i - \bar{y})^2} $$

If the coefficient of determination is close to $1$, then the linear regression model is a good fit for the problem. If it is close to $0$, then the model is not adequate for the problem. Another metric to measure the accuracy of predictions is   the standard error of the estimate (also called the residual standard error or the root mean square error) given by: 

$$ \sigma_{\text{est}} = \displaystyle \sqrt{\displaystyle \frac{\displaystyle\sum_{i=1}^{n}(\hat{y}_i -\bar{y})^2}{n-p}}$$


where $p$ is the number of parameters being estimated. If we calculate the standard error of the estimate for a regression model with $b_0$ and $b_1$, then $p=2$. The smaller is $\sigma_{\text{est}}$ the better is the model because a value close to zero indicates that most observations are close to the fitted line.

Linear regression is a very simple and intuitive method. In addition to its capability to classify data and predict the output of unseen data, it helps to understand the relationship between variables. However, it is only useful when the relationship between variables is linear. It is also sensitive to outliers.

\subsubsection{Logistic Regressions}        

Logistic regression (LogR) is a statistical method of analyzing datasets  in which one or more independent variables (predictors) predict a result quantified with a binary variable. LogR model is based on a  transformation  (\texttt{logit}) which computes the logarithm of the probability of event divided by the probability of no event. It is defined as follows: 

$$\texttt{logit}(p) = \texttt{ln}\displaystyle \frac{p}{1-p}$$ 

where $p = p(y=1|X)$ is the conditional probability of $y$ being true knowing $X$; $X=(x_1, x_2, ..., x_n)$ is the input vector, $y$ is the predicted value, $p$ is the probability of presence of the event, $1-p$ is the probability of absence of the event, and $\displaystyle \frac{p}{1-p}$ are \textit{odds}.

Assuming a linear relationship between the predictors and $\texttt{logit}(p)$, meaning $\texttt{logit}(p)= \beta_0 + \beta_1 x_1 + ... + \beta_n x_n$, we have

$$\displaystyle \frac{p}{1-p} =  e^{\beta_0 + \beta_1 x_1 + ... + \beta_n x_n}$$

$$\text{or}$$ 

$$p = \displaystyle \frac{1}{1 + e^{-(\beta_0 + \beta_1 x_1 + ... + \beta_n x_n)}}$$

Notice that $p$ is exactly the sigmoid function $\sigma$, also known as the logistic function, applied on the weighted inputs. If it returns a value greater than $0.5$, then the event is considered present. If it returns a value less than $0.5$, then the event is  considered absent. If the returned value is close to $0.5$, then the decision regarding the presence of the event is hard to make.

\begin{figure}[h]

\centering
\includegraphics[scale=0.5]{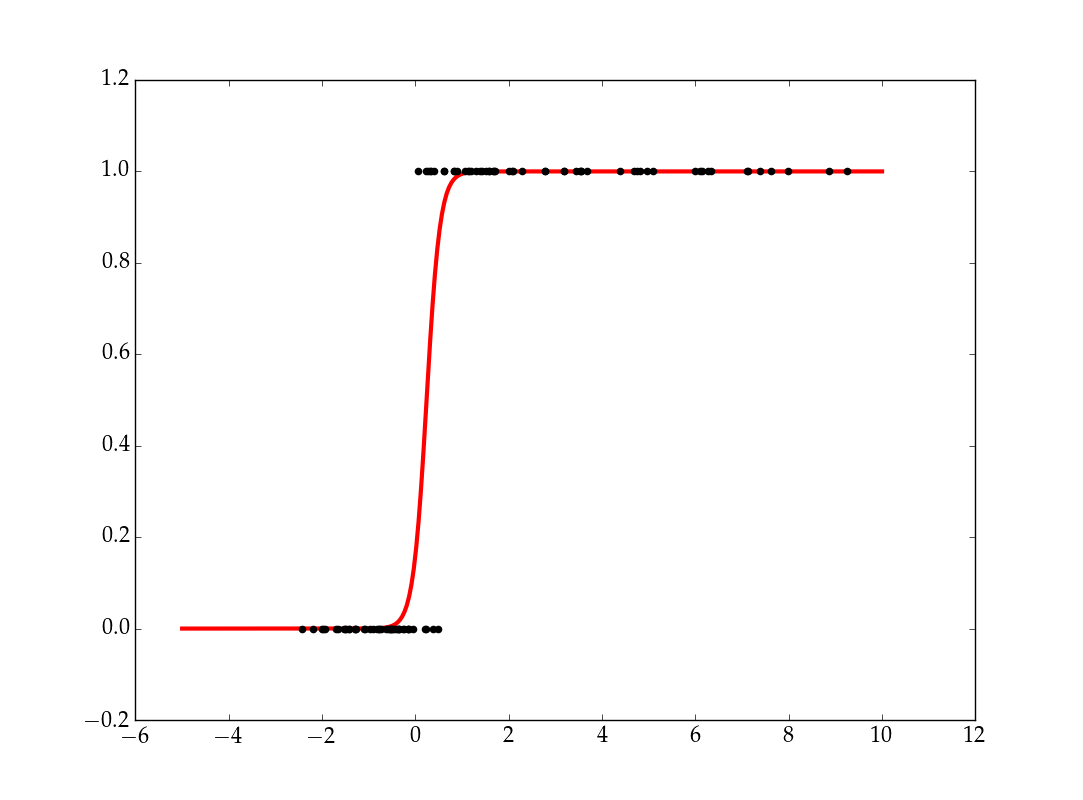}

\caption{Logistic Regression}
\label{fig:LogRegression}
\end{figure}


\begin{figure}[h]
\centering
\includegraphics[scale=0.35,frame]{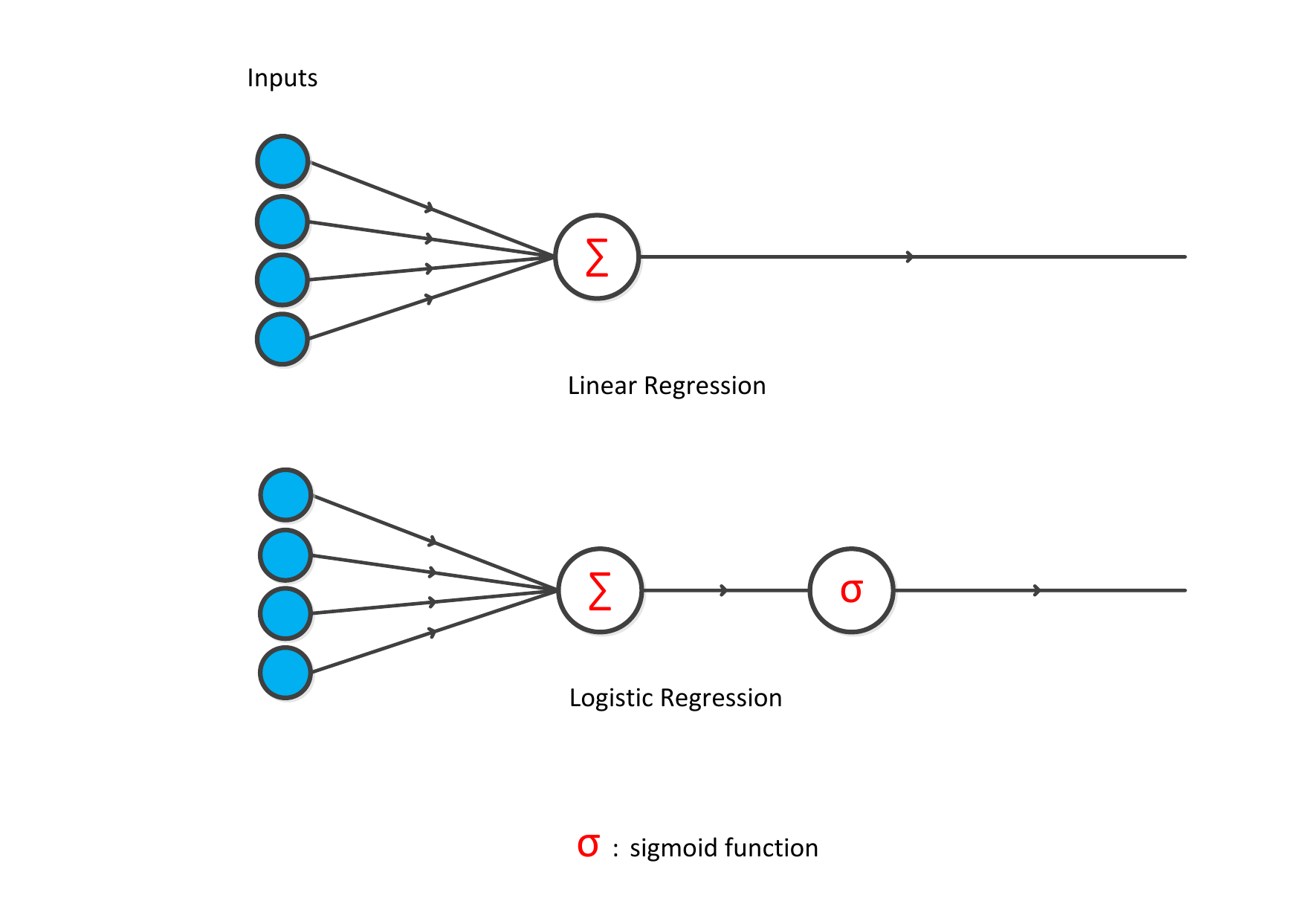}

\caption{Linear Regression versus Logistic Regression}
\label{fig:Regression}
\end{figure}

The learning  process purpose is to determine the best values for the coefficients $\beta_i$. This is reached using the gradient descent. The cost function to be minimized is called the \texttt{LogLoss} given by the following formula

$$ \texttt{LogLoss} = \displaystyle \sum_{(X,y)\in D}^{} -y ~\texttt{ln}(\hat{y}) - (1 - y)  ~ \texttt{ln} (1 - \hat{y})$$

where $D$ is the training dataset, $X=(x_1, x_2, ..., x_n)$ is the input vector, $y$ is the actual labeled value and $\hat{y}$ is  the predicted value. 


In order to remedy the overfitting problem that frequently occurs, a regularization term should be added to the \texttt{LogLoss} function, resulting in the regulated \texttt{RegLogLoss} that should be minimized,  given by the following formula:

$$ \texttt{RegLogLoss} = \displaystyle \sum_{(X,y)\in D}^{} -y ~\texttt{ln}(\hat{y}) - (1 - y)  ~ \texttt{ln} (1 - \hat{y}) + \displaystyle \frac{\lambda}{2} ||w||^2$$

where $\lambda$ is a parameter and $w$ is the weight vector.

Logistic regression is  easy to implement and efficient to train. It provides conditional probabilities which may be very useful in a large number of applications. It is usually possible to transform most non-linear features into linear ones. However, it becomes fairly complicated with multiple predictors.

\subsubsection{Decision Trees }       

\begin{figure}[h]
\centering
\includegraphics[scale=0.35,frame]{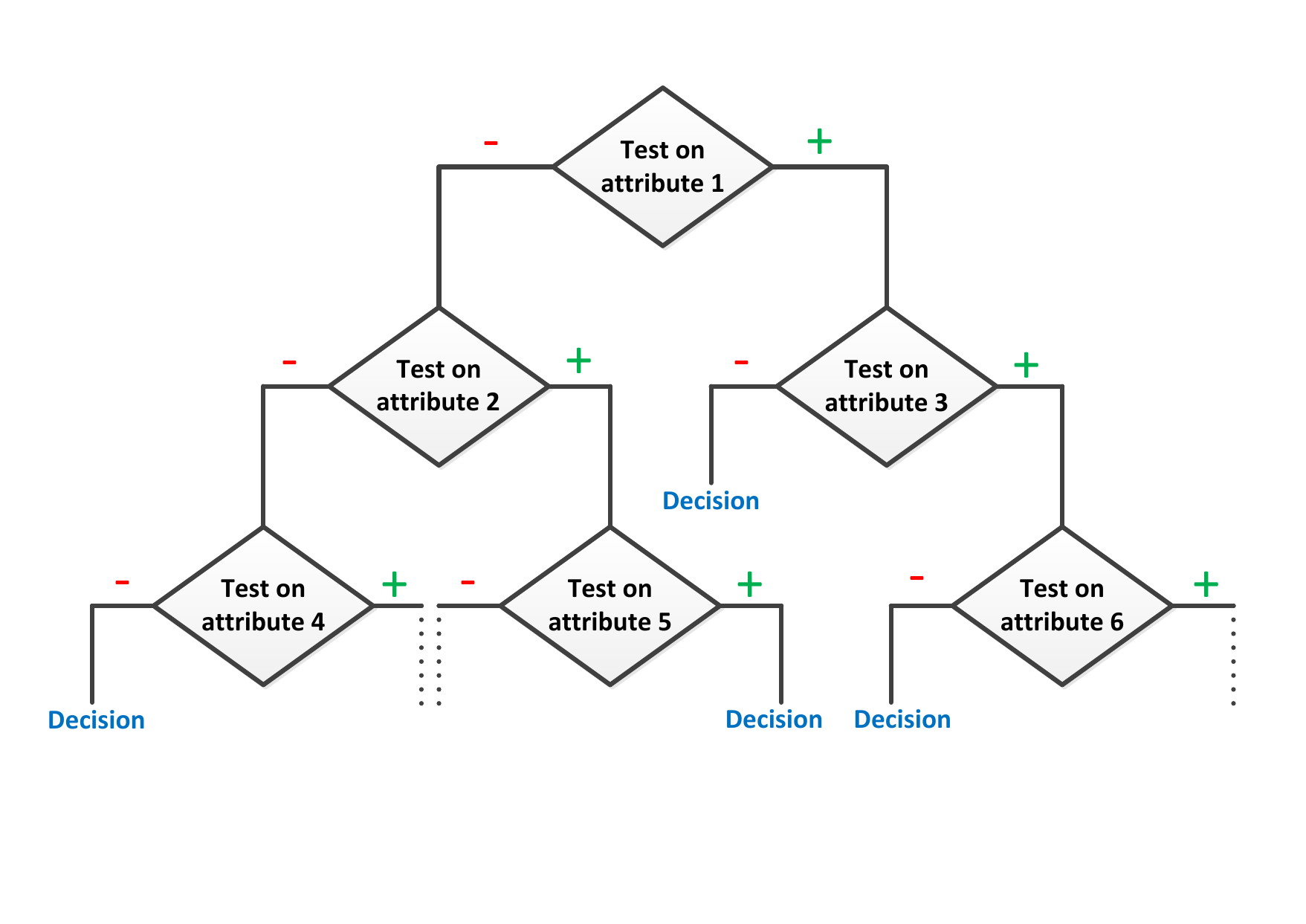}

\caption{Decision tree process}
\label{fig:DT}
\end{figure}

A Decision Tree (DT) classifier is a supervised classification algorithm in Machine Learning. It consists of a tree-like graph where a node is an attribute with a question and a leaf is a class to which belongs an item (see Figure \ref{fig:DT}). It is built as follows. For a training dataset, where each record contains a set of attributes, one of which is the class, a DT classifier attempts to construct a model for the class attribute based on the values of the other attributes. In most cases, a particular dataset  is partitioned into training and testing sets, with the training set serving to construct the model and the testing set serving to validate it. A DT classifies the examples by sorting them from the root to a leaf, and the leaf node gives the answer (the class). Each node in the tree serves as a test case for certain attributes, and each branching edge represents one of the possible answers. This process is run recursively on each subtree rooted at a new node.

For a given dataset, a decision tree is not necessarily unique. For that, we need to find the tree that suits best the dataset, meaning, the tree with the greatest likelihood of correct classification for unseen examples. For that, some algorithms have been suggested such as  the Iterative Dichotomiser 3 algorithm (ID3) \cite{Quinlan1986}. This algorithm builds decision trees using a top-down greedy strategy based on an entropy function and its derived gain function to select attributes in a way that the resulting tree is the smallest possible, meaning, involving as few attributes as possible.  With that in mind, ID3 does not guarantee an optimal tree and can overfit the training data. This algorithm consists of the following steps \cite{WikiID3}:

\begin{enumerate}
\item Compute the entropy and the gain for every attribute. The entropy function is given by: 

$$\mathcal{E}(D) = \displaystyle\sum_{x\in X}^{} -p(x)\texttt{log}_2 (p(x))$$

where $D$ is a given partition, $X$ is the set of classes, and $$\displaystyle p(x)=\frac{ \text{the number of elements of the class }x}{ \text{the number of elements of the data set}}$$

Note that when $S$ is perfectly classified, $\mathcal{E}(S)=0$.


The gain function is given by: 

$$ \texttt{Gain}(D,A) = \mathcal{E}(S) - \displaystyle\sum_{i=1}^{v} \displaystyle \frac{|D_i|}{|D|} \mathcal{E}(D_i)$$

where $D$ is a given partition, $A$ is an attribute having $v$ different values, and $D_i$ is a sub-partition of $D$.
  
\item Split the dataset  into subsets on the attribute having the highest gain once the partitioning is performed;

\item Create a decision tree node using that attribute; and

\item If the dataset  is perfectly classified, then terminate, else, recurse on subsets with the remaining attributes.
\end{enumerate}


The Classification and Regression Trees algorithm (CART \cite{cart93})  supports both regression and classification and  builds binary trees (i.e. a single split results in two children). CART uses a cost function to test all attributes and the one with the smallest cost is selected in each splitting step. For regression problems, the cost function is usually the sum of squares error given by 

$$ SSE = \displaystyle\sum_{}^{}(y_i - \bar{y})^2$$

where $y_i$ is the predictor and $\bar{y}$ is its mean value.

For classification problems, the cost function is the Gini score given by: 

$$ G = \displaystyle\sum_{}^{}(p_i * (1-p_i))^2$$

where $p_i$ is the proportion of same class inputs present in a given group. The best score (i.e. $G=0$) occurs when a group contains only inputs of the same class (i.e. $p_i$ is either $1$ or $0$), and the worst score (i.e. $G=0.5$) occurs when a node has a $50$-$50$ split of classes (i.e. $p_i = 0.5$).

The performance of a tree can be improved by pruning, which consists in getting rid of branches that make use of attributes with low importance.

Other algorithms such as C4.5 \cite{Salzberg1994} and MARS \cite{Friedman91multivariateadaptive} propose other strategies to build decision trees.

Several examples using decision trees can be found in \cite{DTEX1,DTEX2,DTEX3,ijca2017913660}.

Decision tree algorithms are known to generate simple rules that are easy to understand and interpret. They can handle both continuous and categorical features. However,  they are prone to errors in classification problems with multiple classes and can be computationally costly to train. 

\subsubsection{Random Forest}              

Random Forests are Ensemble Learning methods for both classification and regression. They work by building a variety of decision trees during the training phase and provide the predicted class or value by following the general scheme of one of the Ensemble Learning techniques. These techniques are presented in an independent sub-section (see  \ref{secEnsembleLearnining}).


    
\subsection{Naive Bayes Classifiers}

The Naive Bayes classifier is a probabilistic algorithm used to discriminate different entries based on some features. The core of the algorithm is based on Bayes' theorem which posits: 

 $$P(A|B)=\frac{P(B|A).P(A)}{P(B)}$$

where:
\begin{itemize}
\item $A$ and $B$ are two events ($B$ is the evidence and $A$ is the hypothesis);
\item $P(A|B)$ is the posterior probability which is the probability of $A$ happening given that $B$ has happened;
\item $P(B|A)$ (the likelihood) is the posterior probability of $B$ happening given that $A$ has happened;
\item $P(A)$ is the prior probability of $A$ happening; and
\item $P(B)$ is the prior probability of $B$ happening. 
\end{itemize}


The theorem assumes that the features are all independent, which means that the presence of a given feature does not affect another feature. That is why the algorithm is referred to as naive.\\

To understand how the Naive Bayes classifier derives from Bayes theorem, assume a feature vector $X=(x_1, x_2, \mbox{ ... }, x_n)$ and a class variable $y_k$ among $K$ classes in the training data. Bayes's theorem implies: 

\begin{equation} P(y_k|X)=\frac{P(X|y_k).P(y_k)}{P(X)}, k \in \{1,2, \mbox{ ... }, K\}  \label{naive_eq1}\end{equation}

Considering the chain rule \cite{wiki:chain_rule} for multiple events:
\begin{equation} P(A_1  \cap A_{2} \cap \mbox{ ... } \cap A_n)= P(A_1 | A_{2} \cap \mbox{ ... }\cap A_n). P(A_{2} \cap \mbox{ ... } \cap A_n)  \label{naive_eq2}\end{equation}

the likelihood $P(X|y_k)$ could be written:
\begin{equation}  \label{naive_eq3}
\begin{split}
P(X|y_k) & = P(x_1, x_2, \mbox{ ... }, x_n|y_k) \\
 & = P(x_1|x_2, \mbox{ ... }, x_n|y_k).P(x_2|x_3 \mbox{ ... }, x_n|y_k) \mbox{ ... } P(x_{n-1}|x_n|y_k).P(x_n|y_k)
\end{split}
\end{equation}

This is where the independence assumption of Bayes' theorem comes in handy. This assumption states that:

\begin{equation} P(x_i|x_{i+1} \mbox{ ... } x_n| y_k)=P(x_i|y_k), i \in \{1 \mbox{ ... } n\}  \label{naive_eq4}\end{equation}

The likelihood could then be reduced to:

\begin{equation}P(X|y_k) = \prod_{i=1}^{n} P(x_i|y_k) \label{naive_eq5}\end{equation}

The posterior probability $P(y_k|X)$ could also be reduced to: 
\begin{equation} P(y_k|X)=\frac{P(y_k).\prod_{i=1}^{n} P(x_i|y_k)}{P(X)},  k \in \{1,2, \mbox{ ... }, K\}  \label{naive_eq6}\end{equation}

Considering that $P(X)$ is a constant  for all $k \in 1,2,\mbox{ ... }, K$, the Naive Bayes classification problem could be formulated as follows: for the different class values $y_k$, $ k \in \{1,2, \mbox{ ... }, K\} $; maximize 

\begin{equation} P(y_k).\prod_{i=1}^{n} P(x_i|y_k)\label{naive_eq7}\end{equation}

Notice that the probability $P(y_k)$ is the relative frequency of the class $y_k$ in the training data. Also, notice that $P(x_i|y_k)$ could be calculated using known distributions such as the Gaussian (or Normal) distribution\cite{encyclopediaofmath:NormalDistribution}, the Bernoulli distribution \cite{wiki:Bernoulli_distribution}, or the multinomial distribution \cite{wiki:Multinomial_distribution}. In such cases one has Gaussian (or normal) naive Bayes classifier, Bernoulli naive Bayes classifier, or a Multinomial naive Bayes classifier, respectively.

The naive Bayes algorithm is reputed to be computationally fast and simple to implement, and to cope well with classification problems with high dimensional features and limited training datasets. However, it tightly relies on the independence assumption, and consequently  does not cope well with classification problems where this assumption is not satisfied.










\subsubsection{Support Vector Machine}             

\begin{figure}[h]
\centering
\includegraphics[scale=0.6,frame]{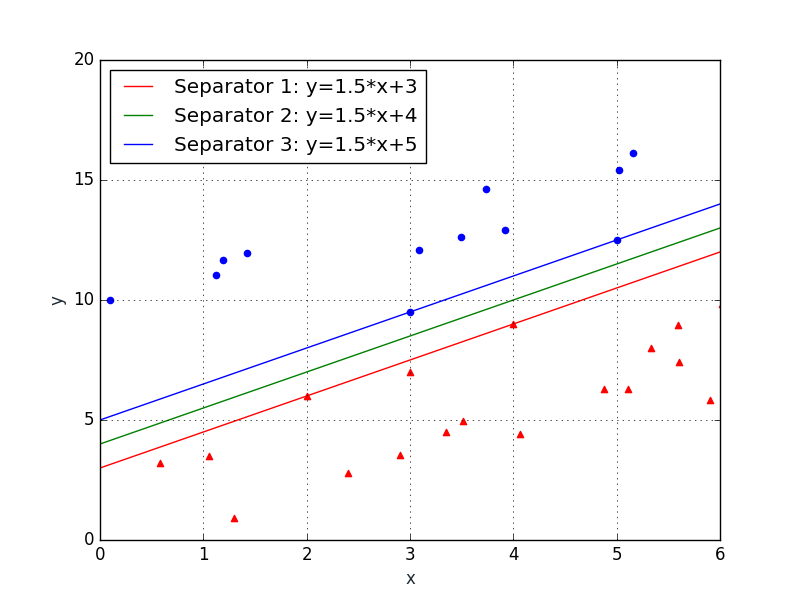}

\caption{Support Vector Machine (in two dimensions)}
\label{fig:SVM1}
\end{figure}

\begin{figure}[!htb]
\centering
\includegraphics[scale=0.6,frame]{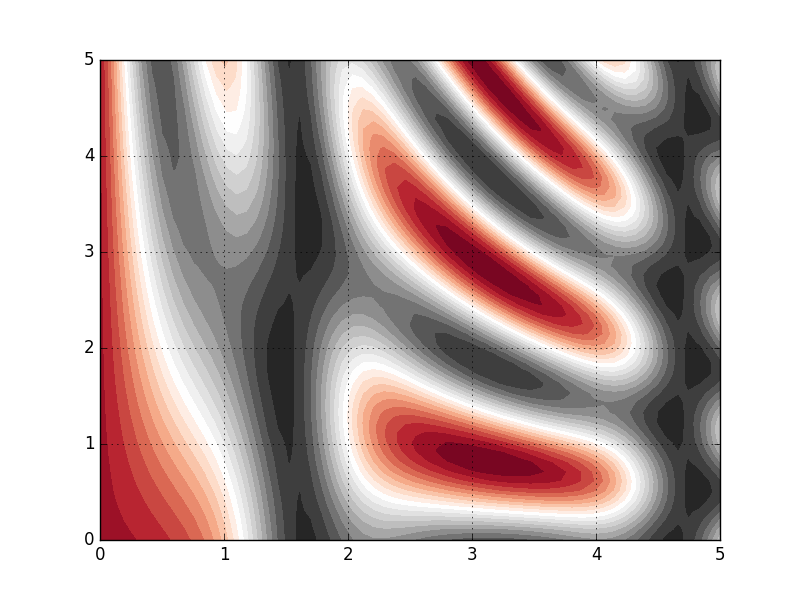}

\caption{SVM (non linearly separable data)}
\label{fig:SVM2}
\end{figure}

\begin{figure}[!htb]
\centering
\includegraphics[scale=0.4,frame]{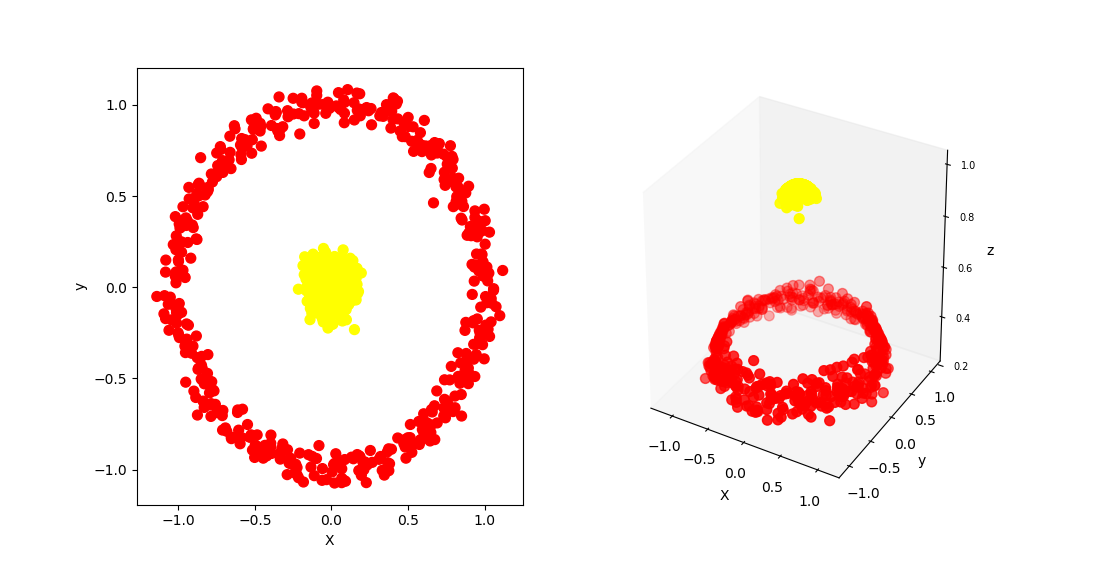}

\caption{SVM (linear separation in a higher-dimensional space using kernels)}

\label{fig:SVM3}
\end{figure}

A Support Vector Machine (SVM) algorithm is a Machine Learning algorithm used to solve classification, regression, and anomaly detection problems. It is known for its solid theoretical foundation.  Its purpose is to separate data into classes using a boundary in such a way that the distance between different data classes and the boundary is as large as possible. This distance is also called \textit{margin} and the SVM is referred to as \textit{wide margin separator}. The \textit{support vectors} are generated by the data points closest to the boundary. Support vectors are the critical elements of the training set and if they change, this will probably change the position of the separating line.  In Figure \ref{fig:SVM1}, in a two-dimensional space, the boundary is most likely the green line, the support vectors are most likely determined by the two points on the blue line as well as the two points on the red line, and the margin is most likely the distance between the boundary and the blue and red lines. Margin maximization ensures robustness against noise and allows the model to be more generalizable. The basic concept of linear boundary implies that the data should be linearly separable, which is rarely the case (see Figure \ref{fig:SVM2}). To overcome this shortcoming, SVMs often use what are called \textit{kernels}\cite{kernel1} which are mathematical functions that allow SVM to perform a linear separation of a non-linearly separable problem in a higher-dimensional space  (see Figure \ref{fig:SVM3}). The theory underpinning SVM is quite challenging and introduces several advanced mathematical concepts. We will devote an annex to cover most of these aspects.

SVM has many advantages. Depending on the data, the performance of SVMs is at least of the same order as that of neural networks. It is also a robust technique because it determines the optimal hyperplane boundaries using the nearest points (support vectors) only, not distant points. However, SVM is reputed to be an opaque technique and very sensitive to the choice of kernel parameters. In addition, its  computation time is often long during the learning phase.







\newpage
\subsubsection{k-Nearest Neighbors}       

$k$-Nearest Neighbors ($k$-NN), where $k$ is an integer, is a very simple and intuitive classification algorithm. Its purpose is to classify target points  of unknown classes according to their distances from $k$ points of a learning sample  whose classes are known in advance. $k$-NN is so a supervised algorithm.  Each point $P$ of the sample is considered as a vector of $\bm{\mathbb{R}}^n$ described by its coordinates $(p_1,p_2, ..., p_n)$. In order to determine the class of a target point $Q$, each of the closest  $k$ points  to it makes a vote. The class of $Q$ corresponds to the class having the majority of votes. $k$-NN could use many metrics in a normed vector space to find the closest points to $Q$, for instance:

\begin{itemize}
\item the Euclidean distance:$$d(P,Q)= \sqrt{\displaystyle\sum_{i=1}^{n} (p_i - q_i)^2}$$

\item the Manhattan distance, defined by: $$d(P,Q)= {\displaystyle\sum_{i=1}^{n} |p_i - q_i|}$$

\item the Minkowski distance, defined by $$d(P,Q)= \sqrt[j]{\displaystyle\sum_{i=1}^{n} |p_i - q_i|^j}$$ where $j$ is an integer number such that $j \ge 1$. It is a  generalization of both the Euclidean distance ($j=2$) and the Manhattan distance ($j=1$).

\item the Tchebychev distance, defined by: $$d(P,Q)= \displaystyle \max_{i\in\{0, ..., n\}}(|p_i - q_i|)$$

\item the Canberra distance, defined by: $$d(P,Q)=  {\displaystyle\sum_{i=1}^{n} \frac{|p_i - q_i|} {|p_i|+|q_i|} } $$

\end{itemize}

The value of $k$ influences both the accuracy and the performance of the algorithm.  That is, a small value of $k$ implies that the noise may have a significant impact on the result and a large value makes the calculation expensive. The $k$ value is generally determined based on the empirical results. $ $

The $k$-NN algorithm has many advantages. First, it is simple, intuitive, and well-performing with small datasets. In addition, it makes no particular assumption and does not actually build any model. It also constantly evolves  as new training data is collected. However, $k$-NN becomes slow as the dataset   grows. It also does not contend well with imbalanced data. That is, if we have two classes, \textit{Normal} and \textit{Abnormal}, and a predominate number of the training data labeled as \textit{Normal}, then the algorithm strongly prefer to assign individual data points to the \textit{Normal} class instead of the \textit{Abormal} one. The $k$-NN algorithm does not deal with missing values as well.





\subsubsection{Artificial Neural Networks}       
\begin{figure}[h]
\centering
\includegraphics[scale=0.4,frame]{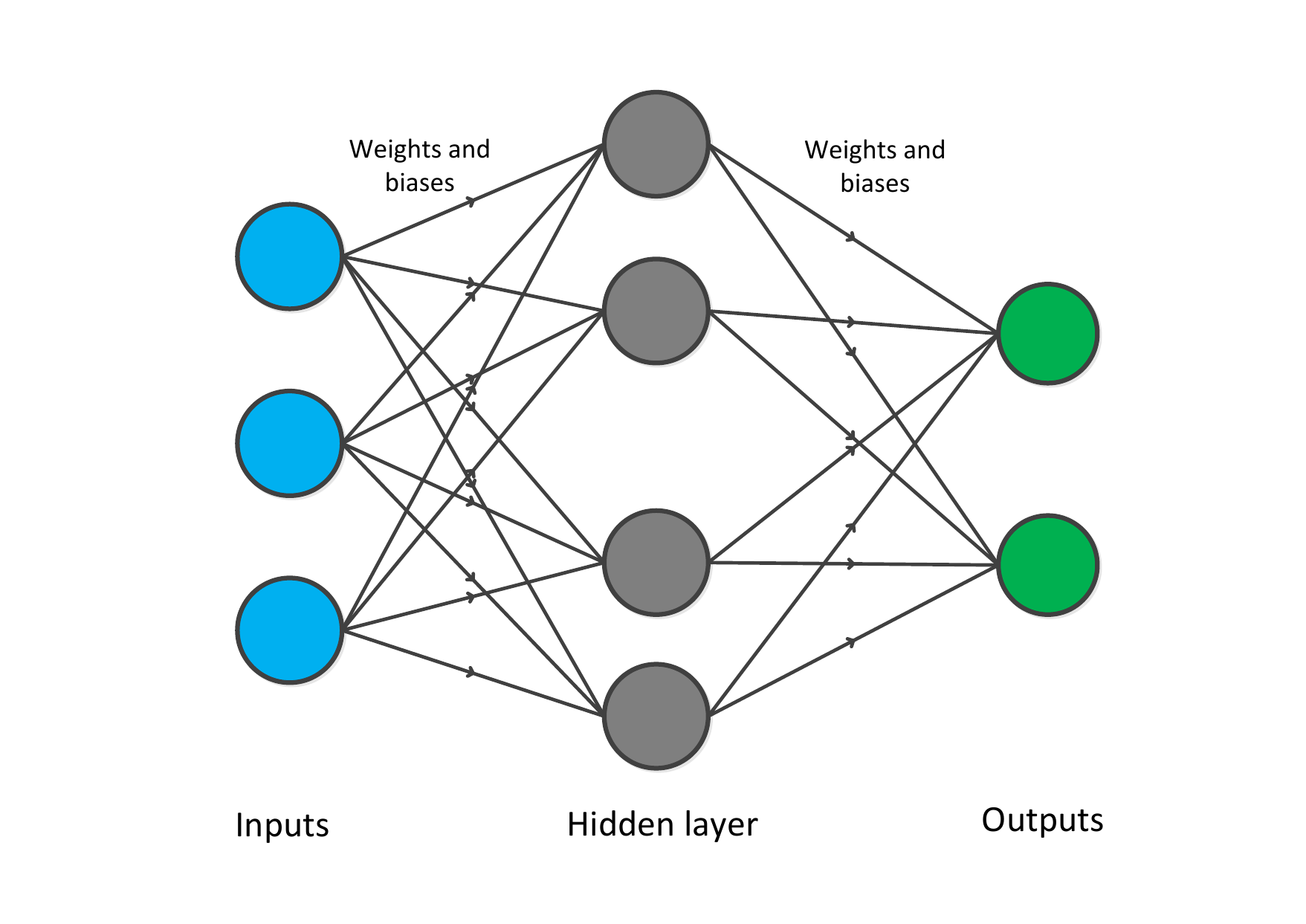}

\caption{Artificial Neural Network}
\label{fig:ANN}
\end{figure}

An Artificial Neural Network (ANN) is a layered model. It is widely used in artificial intelligence. Its structure is very similar to  the network of neurons in human brain with layers of linked nodes. A neural network can be trained so that it can recognize patterns, classify data, and predict future outcomes. An artificial neural network, as given by Figure \ref{fig:ANN}, consists of the following layers and components:

\begin{itemize}
\item an input layer, usually denoted by $x$;
\item a hidden layer, with a set of neurons;
\item an output layer, usually denoted by $\hat{y}$;
\item connections between all the input layer nodes and the hidden layer nodes, and between the hidden layer nodes and the output layer nodes;
\item a set of weights and biases for each connection; and
\item an activation function.
\end{itemize}

\begin{figure}[h]
\centering
\includegraphics[scale=0.4, frame]{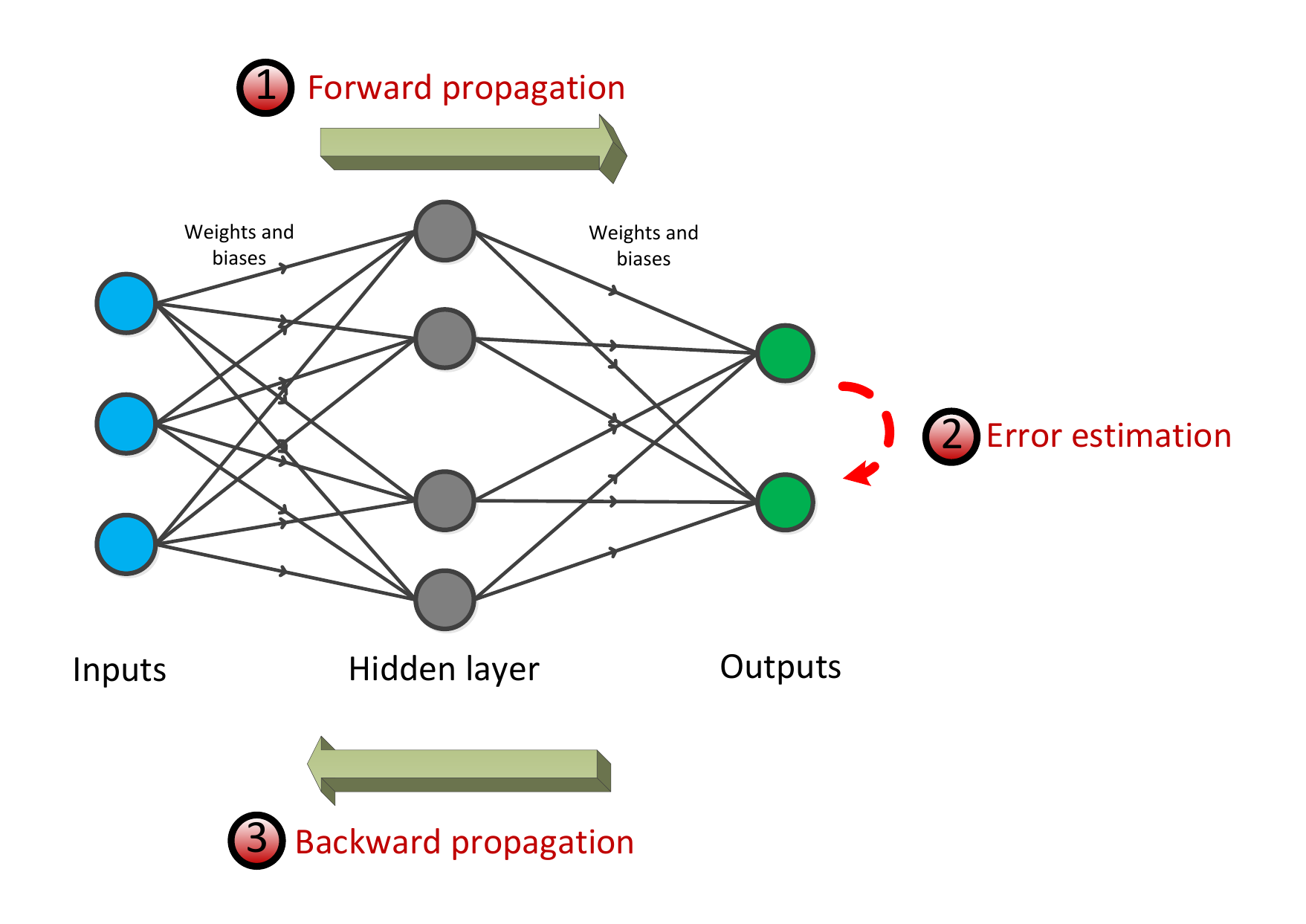}

\caption{ANN: forward and backward propagation}
\label{fig:BackPropagation}
\end{figure}

The training process of an ANN consists in the steps shown in Figure \ref{fig:BackPropagation}.

\paragraph{Forward propagation}

The goal of this step is to calculate the output vector $\hat{y}$ from an input vector $x$. During this step, weights and biases are decided and an activation function is applied to the resulting output. For a basic neural the output is given by the following equation:

$$\hat{y} = \sigma(W_2\sigma(W_1x +b_1)+b_2)$$

where $\sigma$ is an activation function, $W_1$ and $W_2$ are weights, and $b_1$ and $b_2$ are biases. In this step, the information flows in only one direction, from the input layer to the output layer.

\paragraph{Error estimation}

Once the forward propagation is done, one needs to evaluate how \textit{good} the prediction is. For that, a loss function (also called cost function or objective function) is used to estimate the error between the ground truth $y$ and the predicted output $\hat{y}$. Various loss functions \cite{ZhangHao} can be used for that purpose and their performance depends on the nature of the problem. The most common loss functions are given in Table \ref{tabLoss}.

\begin{table}
\centering
\begin{tabular}{|L{5cm}|L{5cm}|}
\hline
\ZZ \textbf{Loss function name} &    \textbf{Equation} \\
\hline
 Mean squared error (MSE)&    $\displaystyle \frac{1}{n}\displaystyle\sum_{i=1}^{n} (y_i - \hat{y}_i)^2$  \\\hline

 Mean squared logarithmic error (MSLE) &    $\displaystyle \frac{1}{n}\displaystyle\sum_{i=1}^{n} ( \texttt{ln}(y_i +1) - \texttt{ln}(\hat{y}_i+1))^2$  \\\hline

 L2 &    $\displaystyle\sum_{i=1}^{n} (y_i - \hat{y}_i)^2$  \\\hline

 Mean absolute error (MAE)&    $\displaystyle \frac{1}{n}\displaystyle\sum_{i=1}^{n} |y_i - \hat{y}_i|$  \\\hline

 Mean absolute percentage error (MAPE)&    $\displaystyle \frac{1}{n}\displaystyle\sum_{i=1}^{n} |\displaystyle \frac{y_i - \hat{y}_i}{y_i}|.100$  \\\hline

 L1 &    $\displaystyle\sum_{i=1}^{n} |y_i - \hat{y}_i|$  \\\hline

 Cross entropy (CE)&    $\displaystyle \frac{1}{n}\displaystyle\sum_{i=1}^{n} [y_i \texttt{ln} (\hat{y}) + (1-y_i)\texttt{ln}(1-\hat{y}_i)]$  \\\hline

 Negative logarithmic likelihood (NLL)&    $-\displaystyle \frac{1}{n}\displaystyle\sum_{i=1}^{n} \texttt{ln}(\hat{y}_i)$  \\\hline

 Poisson &    $-\displaystyle \frac{1}{n}\displaystyle\sum_{i=1}^{n} (\hat{y}_i - y_i.\texttt{ln}(\hat{y}_i))$  \\\hline

 Cosine proximity &    $-\displaystyle \frac{y.\hat{y}}{||y|| || \hat{y}||}$; "$.$" is the scalar product  \\\hline

 Hinge &    $\displaystyle \frac{1}{n}\displaystyle\sum_{i=1}^{n} \texttt{max} (0, 1- y_i.\hat{y}_i)$  \\\hline

 Squared Hinge &    $\displaystyle \frac{1}{n}\displaystyle\sum_{i=1}^{n} (\texttt{max} (0, 1- y_i.\hat{y}_i))^2$  \\\hline

\end{tabular}
\caption{Common loss functions}
\label{tabLoss}
\end{table}

\paragraph{Backward propagation} 

\begin{figure}
\centering
\includegraphics[scale=0.5]{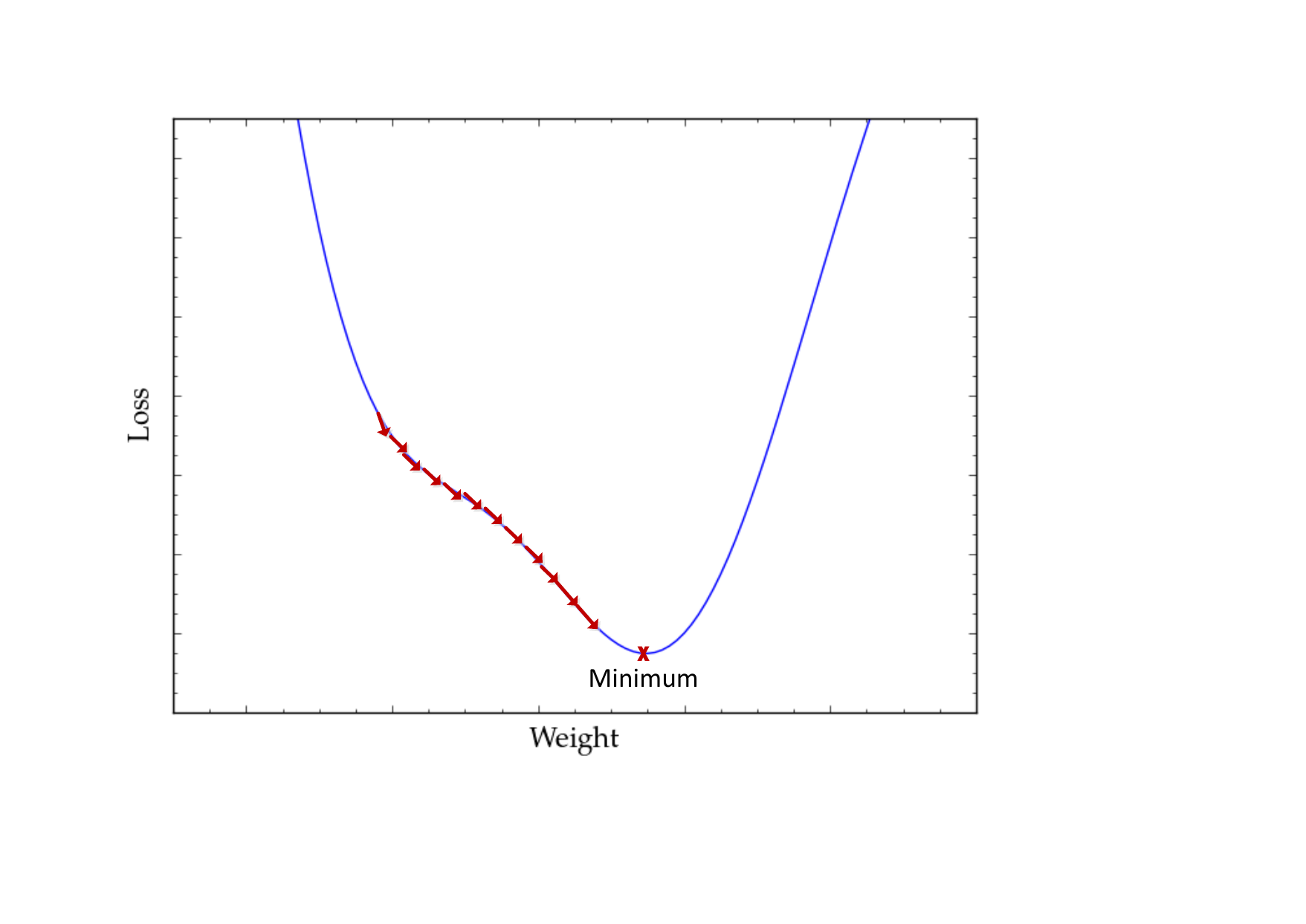}
\caption{Gradient descent}
\label{fig:Gradient}
\end{figure}

Once the error is measured by a loss function, one has to define the strategy to propagate it backward in order to adjust weights and biases in a manner that makes the network converge towards the optimal output. The optimal output obviously corresponds to the point where the loss function reaches its minimum. For that, one needs to calculate the gradient of the loss function. The gradient indicates in which direction the network should walk to get closer to the minimum (i.e. the tangent direction given by partial derivatives). This step is called the gradient descent (see Figure \ref{fig:Gradient}). 

For a loss function $\mathcal{L}$, weights and biases are updated according to the following iterative equation \cite{ChiFengWang}:

$$
\left\{
    \begin{array}{l}
        W_{t+1} =W_t - \eta \displaystyle\frac{\partial \mathcal{L}}{\partial W} \\
\\
        b_{t+1} = b_t - \eta \displaystyle\frac{\partial \mathcal{L}}{\partial b}
    \end{array}
\right.
$$

\begin{figure}
\centering
\includegraphics[scale=0.5]{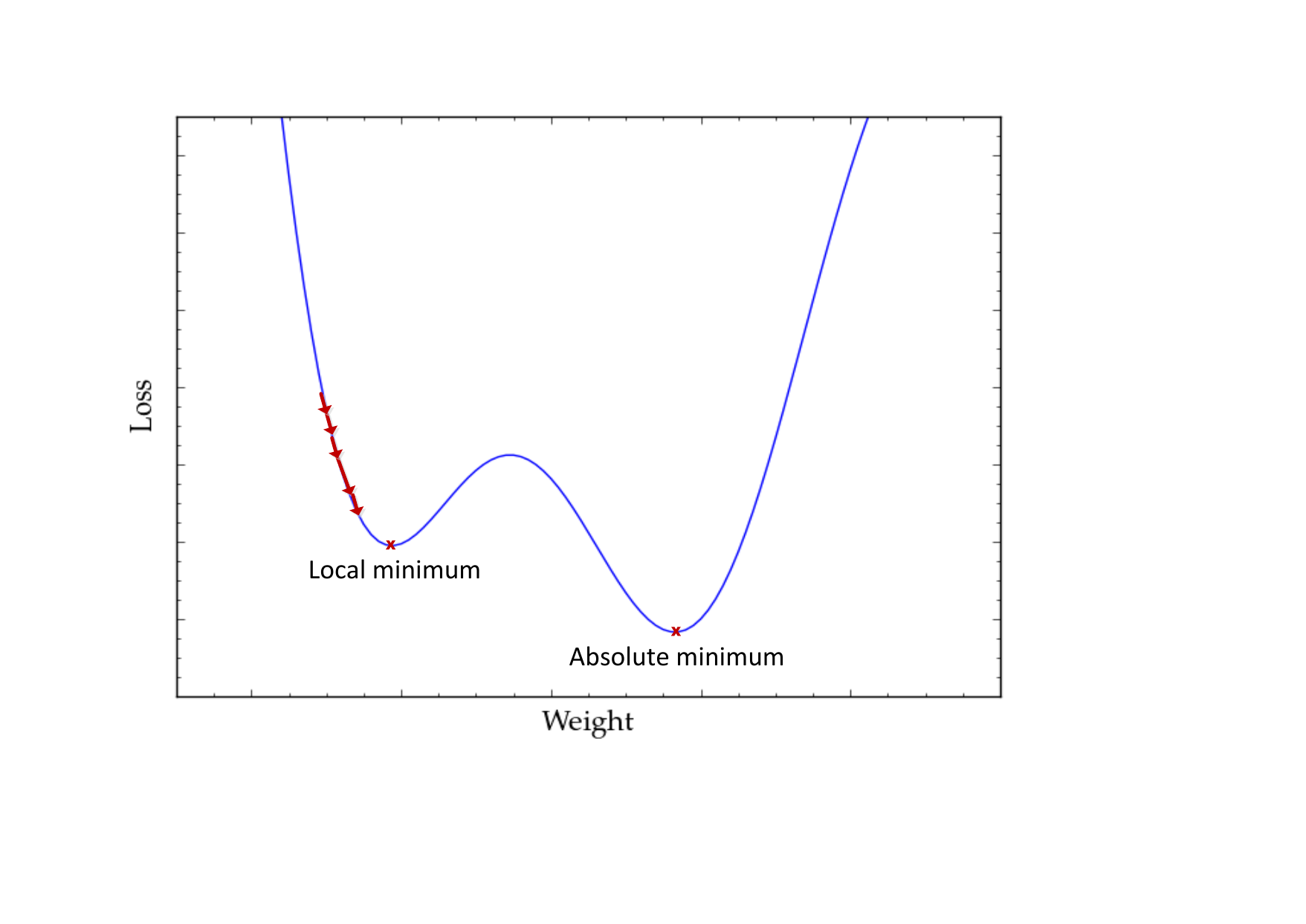}
\caption{Convergence to a local minimum}
\label{fig:LocalMinimum}
\end{figure}

The factor $\eta$ is called the  learning rate. Its role is to determine to what extent the weights and biases should move towards the minimum. The larger $\eta$ is, the bigger is the correction at each step. If it is set too large, the descent can jump over the minimum, which leads to oscillations around it or even to a clear divergence. If it is set too small,  the convergence to the bottom is sure but may take a very long time. This being said, in some cases the descent may lead to a local minimum instead of the absolute one as shown in Figure \ref{fig:LocalMinimum}.

Gradient descent may be costly in terms of convergence speed. Optimizers have been developed to achieve a trade-off between convergence speed and accuracy. In this document, we reserve an annex to discuss these optimizers including their mathematical foundation.

Forward propagation, error estimation, and backward propagation are repeated until a fairly small loss is reached. The initial weights and biases are given randomly.











\newpage

 \section{Deep Learning techniques}

\subsection{Deep Learning}        

Deep Learning (DL) is a category of algorithms deriving from artificial neural networks and inspired by the human brain behaviour. The network is made up of tens or even hundreds of layers of neurons, each receiving and interpreting the information from the previous layer. At each step, the wrong answers are returned to the upstream levels to correct the mathematical model. As the program progresses, it reorganizes the information into more complex blocks. When the model is subsequently applied to other cases, it is able to classify data that it has never experienced.







\subsection{Deep Learning Algorithms}        

\vspace{0.5cm}

\subsubsection{Multilayer perceptron}       

\begin{figure}[!htb]
\centering
\includegraphics[scale=0.4,frame]{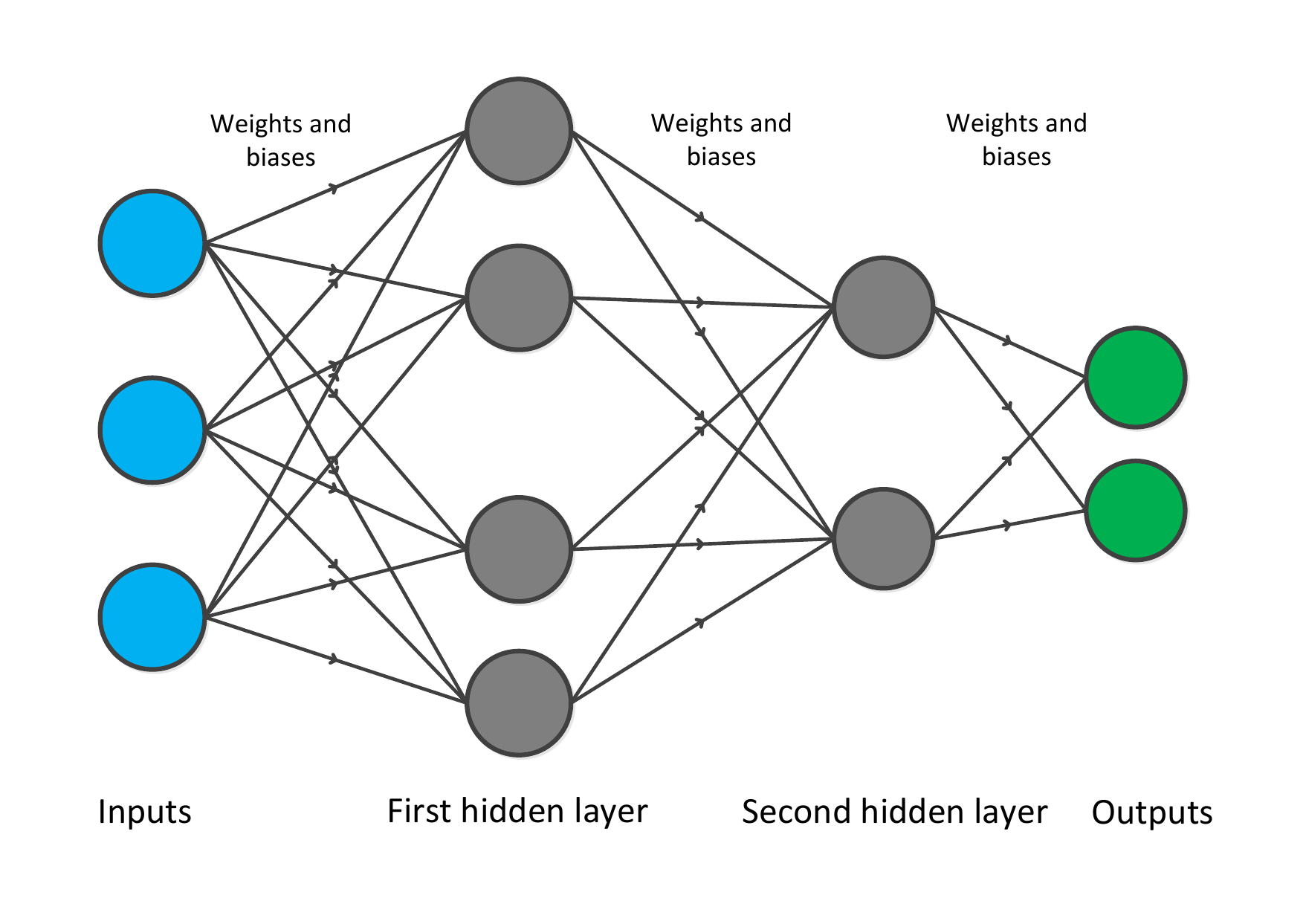}

\caption{Multilayer perceptron (two hidden layers)}
\label{fig:MultiPerc}
\end{figure}

The multilayer perceptron, as given by Figure  \ref{fig:MultiPerc}, is organized in three parts

\begin{enumerate}

\item The input layer: a set of neurons that carry the input signal;

\item The hidden layers: they constitute the heart of the perceptron. This is where the relationships between the variables will be established. Choosing the right number of neurons per layer and the right number of layers is a difficult problem. However, in general, a small number of layers is sufficient for most problems and a large number of layers often leads to overfitting problems. In practice, there are at least as many neurons per layer as there are of inputs, but this is not always the case; and

\item The output layer: this layer represents the final result of the network (the prediction). 

\end{enumerate}

\pagebreak
\subsubsection{Recurring Neural Networks}       

\begin{figure}[!htb]
\centering
\includegraphics[scale=0.4]{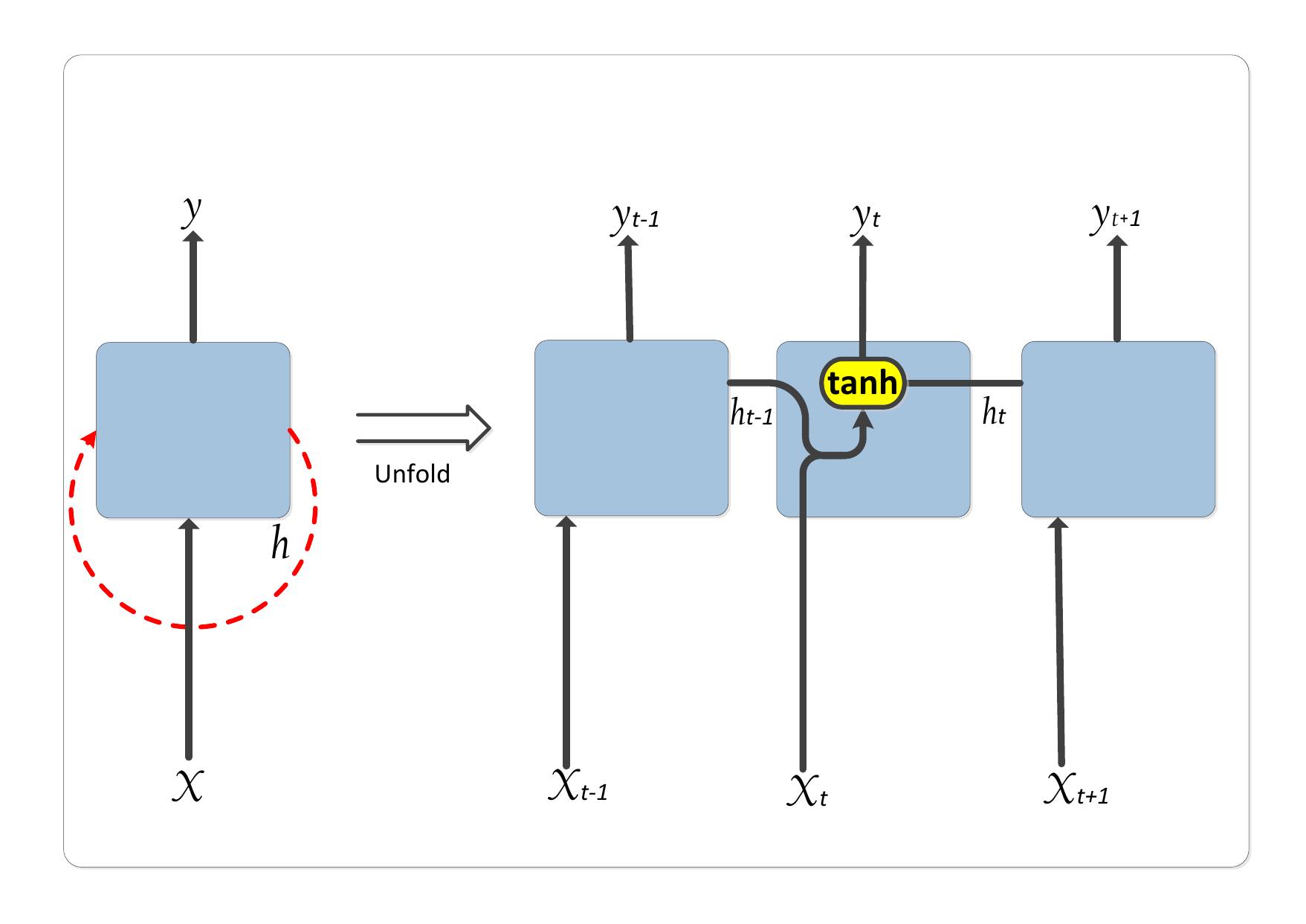}

\caption{RNN (repeating module)}
\label{RNN}
\end{figure}

Recurring Neural Networks (RNNs) take into account not only the current input they see, but also the input they have previously perceived over time. That means that the decision made at the time step $t-1$ affects the decision that will be made at the upcoming time step $t$. Hence, an RNN has two sources of input, the present and the recent past. Adding memory to neural networks has a goal. In fact, the sequence itself is information that should be taken care of. The sequential information is maintained in an internal memory (hidden state(s)). This latter is crucial to capture long-term dependencies between events that are distant in time. A simple RNN (see Figure \ref{RNN}) has a short-term memory, a single layer and a recurring connection with itself that allows it to learn the impact of the previous input $x_{t-1}$ as well as the current input $x_t$ on the prediction $y_t$. It gives the RNN a contextual sense. At a time $t$, the hidden state $h_{t-1}$ is fed into the network and merged with the input $x_t$ to provide the hidden states $h_t$ and $y_t$. During the back-propagation time, the RNN learns to tweak different weights to reflect the repercussion of the current input and the previous output on the current output.  A non-linear activation function, usually the hyperbolic tangent function, is used to squash the values of the hidden states $h_t$ and $y_t$ at every step. It is acceptable to assume that $h_0$ is a vector of zeros. The equations of the hidden states are as follows:

\begin{center}
\begin{tabular}{lllll}
 $h_t $& $=$  &   $\texttt{tanh}(W_x x_t + U_h h_{t-1} + b_h)$ \\ \\
$y_t$ &$=$ &   $\texttt{tanh}(W_y h_t + b_y)$\\
\end{tabular}

\end{center}

where $W_x, U_h \mbox{ and }W_h$, are weights and $b_h \mbox{ and } b_y$ are biases, all adjusted during the back-propagation time. More complex RNNs are also available such as one-to-many, many-to-one and many-to-many RNNs \cite{Karpathy} as well as Jordan networks \cite{JORDAN1997471}. 

Although RNNs cope efficiently  with sequences, especially short sequences, they may suffer from a number of problems \cite{Bengio}. The first one is the exploding problem. This may appear when the model weights quickly become very large during the training process. The second problem is the vanishing gradient when the values of the gradient become extremely small, this causes the model to cease learning. This may restrict RNNs from processing long sequences (e.g. long paragraphs) with efficacy. In addition, RNNs are reputed to be sensitive to how the model is initialized. 
This may lead to a model-overfitting problem and hence strange results could be encountered when an RNN is applied to data that the model has not seen during its learning phase. This being said, it is worth mentioning some recent research on a breed of RNNs called Independently RNNs (IndRNNs) \cite{IndRNNs}, claiming that it is possible to address the RNN limits in an effective manner. However, this requires further confirmation from independent researchers and industrialists since these models are not yet sufficiently studied and experimented with.


\pagebreak

\subsubsection{Long Short-Term Memory Networks}       

\begin{figure}[!htb]
\centering
\includegraphics[scale=0.4]{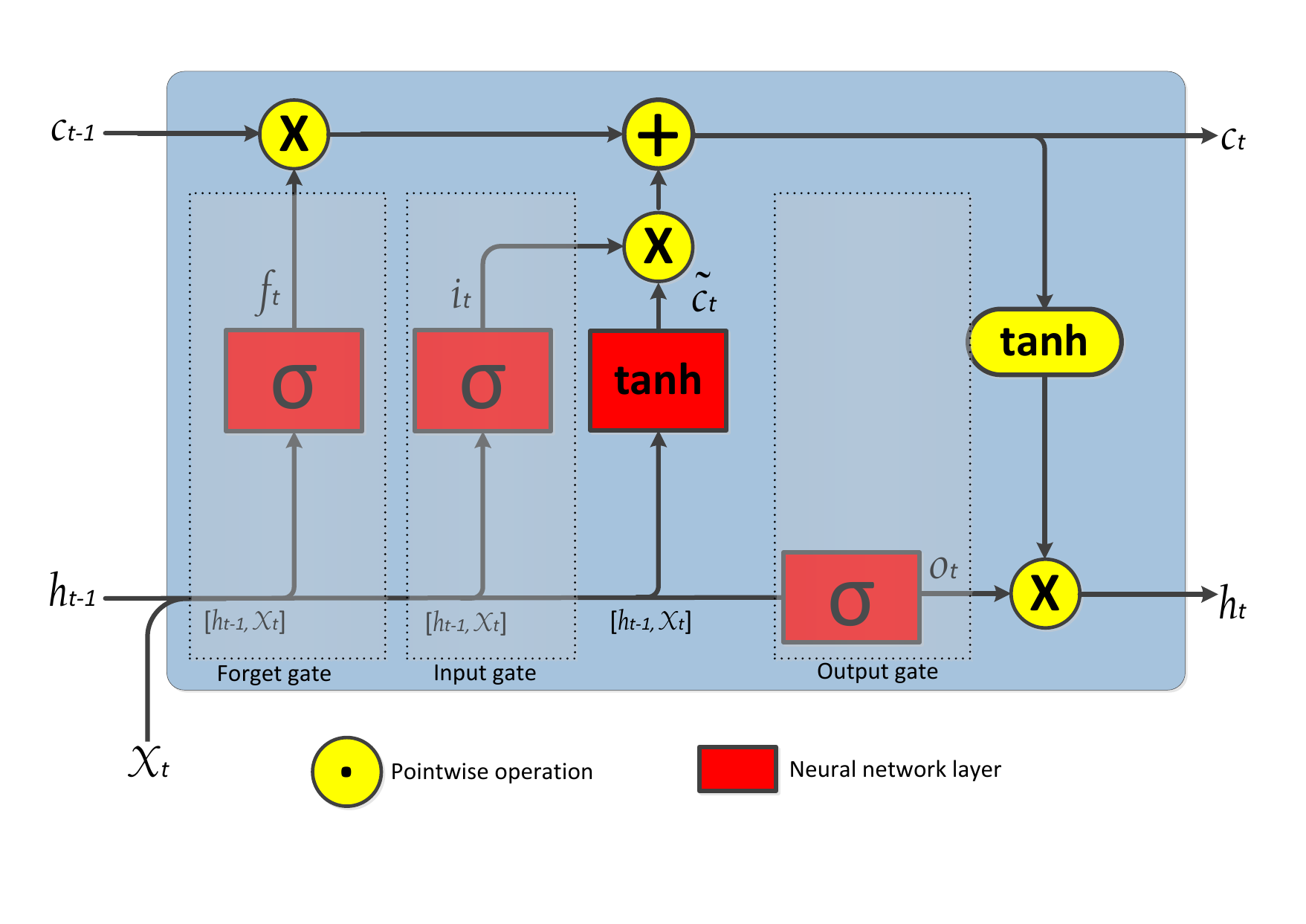}

\caption{LSTM (cell)}
\label{LSTM}
\end{figure}

Long Short-Term Memory  (LSTM)  networks \cite{hochreiter1997long} pertain to the family of RNNs in the sense that they are able to learn long-term dependencies and deal with sequences. They were devised to overcome the RNN  problems mentioned above. The key element of an LSTM is its cell state which acts as the memory of the network that carries the relevant information during the sequence processing. In simple RNNs, the repeating module consists of a single \texttt{tanh} layer. In LSTMs, the repeating module is more complex and consists of three components called gates (see Figure \ref{LSTM}). The first gate is called a forget gate, which consists of a neural network layer with sigmoid. 
At a time step $t$, this layer examines the previous hidden output $h_{t-1}$ as well as the current input $x_t$. These two inputs are first concatenated, then multiplied by weight matrices and a bias is added. Then the sigmoid function is applied. This is expressed by the equation: 

\begin{center}
\begin{tabular}{r c l }
 $f_t$ & $=$ & $\sigma(W_f . [h_{t-1},x_t] + b_f) $
\end{tabular}
\end{center}

The sigmoid function generates a vector with values ranging from $0$ to $1$.  Basically, its role is to decide which values the cell state should maintain and which values should be thrown away. If the sigmoid outputs $0$ for a given element in the vector, that means that the forget gate wants the cell state to forget it. If it outputs $1$,  that means that the forget gate want the cell state to maintain it. This vector is then multiplied (pointwise multiplication) \cite{wiki:pointwise} by the previous cell state $c_{t-1}$. 

The second gate is called the input gate. Its role is to determine what new information should be added to the cell state. This is made possible thanks to two layers. The first one is a sigmoid that outputs a vector of $0$ and $1$ values that determine what should be modified and what should not be. This is illustrated by the equation: 

\begin{center}
\begin{tabular}{r c l }
 $i_t$ & $=$ &  $\sigma(W_i . [h_{t-1},x_t] + b_i) $
\end{tabular}
\end{center}

The second layer is a \texttt{tanh} layer that creates a vector of new candidate values $\tilde{c}$ that could be added to the  cell state. This is expressed by the equation: 

\begin{center}
\begin{tabular}{r c l }
 $\tilde{c}_t$ & $=$ & \texttt{tanh}$(W_c . [h_{t-1},x_t] + b_c)$  
\end{tabular}
\end{center}

The \texttt{tanh} function is used because its output can be positive or negative, allowing for both increases and decreases in the cell state.

Now, it is time to update the cell state considering the two previous gates' outputs. This is expressed by the equation 

\begin{center}
\begin{tabular}{r c l }
 $c_t$ & $=$&  $(f_t \times c_{t-1}) + (i_t \times \tilde{c}_t)$ 
\end{tabular}
\end{center} 

The third gate is the output gate. This gate decides what the cell will output. The output depends on a filtered version of the cell state. For that, a sigmoid layer, similar to the two sigmoid layers of the previous gates, will first decide which parts of the cell state will be output by generating a vector of $0$ and $1$ values. Then, a \texttt{tanh} function squashes the updated cell state. Finally, a pointwise multiplication is carried out between the sigmoid output vector and the filtered cell state. The output is the hidden output state $h_t$. This is expressed by the following equations: 

\begin{center}
\begin{tabular}{r c l }
 $o_t$ & $=$ &  $\sigma(W_o . [h_{t-1},x_t] + b_o) $   \\   \\ 
 $h_t$ & $=$& $o_t \times$ \texttt{tanh}$(c_t)$  
\end{tabular}
\end{center}

There are some variations of LSTM. For instance, Gers et Schmidhuber  \cite{Gers2000RNT870462} added connections, called peephole connections, between gates and the cell state  to ameliorate the learning of long term dependencies and to deal with longer sequences. These gates have the following equations:
\begin{center}
\begin{tabular}{r c l }
 $f_t$ & $=$ & $\sigma(W_f . [c_{t-1},h_{t-1},x_t] + b_f) $  \\   \\ 
 $i_t$ & $=$ &  $\sigma(W_i . [c_{t-1},h_{t-1},x_t] + b_i) $  \\   \\ 
 $o_t$ & $=$ &  $\sigma(W_o . [c_{t-1},h_{t-1},x_t] + b_o) $   \\   \\ 
\end{tabular}
\end{center}

Other LSTM variations are summarized in Table \ref{tabLSTMVAR}.

\pagebreak

\begin{table}[!htb]

\centering

\begin{tabular}{|c|c|}
\hline
\ZZ \textbf{LSTM variation} &   \textbf{Description}   \\ 
\hline
\ZZ NIG &    No input gate (i.e. $i_t = 1$)    \\\hline
\ZZ NFG &  No forget gate  (i.e. $f_t = 1$)   \\\hline
\ZZ NOG &  No output gate  (i.e. $o_t = 1$)   \\\hline
\ZZ NIAF &  No input activation function   \\\hline
\ZZ NOAF &  No output activation function    \\\hline
\ZZ CIFG &  Coupled input and forget gate (i.e. $f_t = 1 - i_t$)   \\\hline
\ZZ NPH &  No peepholes    \\\hline
\ZZ FGR &  Full gate recurrence   \\\hline
\end{tabular}

\caption{ LSTM variations}
\label{tabLSTMVAR}
\end{table}

A comprehensive discussion of these LSTM variations' results is available in \cite{GreffSKSS15}.\\

LSTMs are better than RNNs in remembering information for a long time since they have the capability of removing or adding information to their cell states thanks to their gates. However, they are more opaque and harder to debug. They are also more time-consuming than RNNs owing to their additional built-in neural network layers.




\subsubsection{Gated Recurrent Units}       

\begin{figure}[h]
\centering
\includegraphics[scale=0.4]{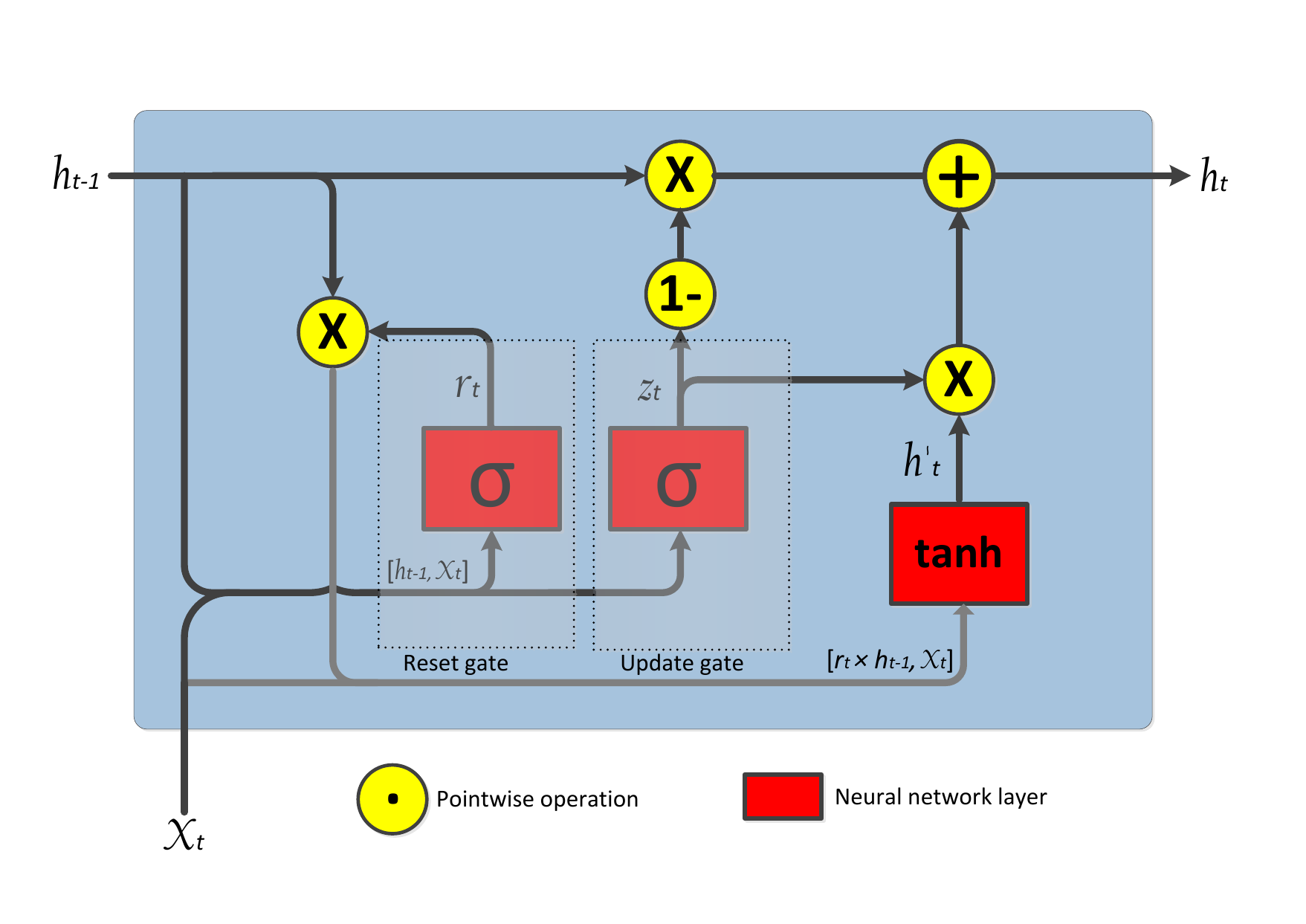}

\caption{GRU (cell)}
\label{GRU}
\end{figure}

Gated Recurrent Unit (GRU) networks \cite{GRUchoetal} also belong to the RNN family. They are also designed to overcome their limitations in particular the vanishing gradient problem. There are similarities and differences between GRU and LSTM. The main similarity is that both use gates. The main differences are that GRU does not contain a cell state variable and contains fewer gates than the LSTM. Figure \ref{GRU} describes a GRU cell.\\

Basically,  a GRU cell has two gates: reset and update gates. It takes in entry the previous hidden output $h_{t-1}$ and the current input $x_t$, concatenates them, multiplies them by weights and applies a sigmoid function to squeeze the output between 0 and 1. This is given by the equation:

\begin{center}
\begin{tabular}{r c l }

 $r_t$ & $=$ &  $\sigma(W_r . [h_{t-1},x_t]) $\\  

\end{tabular}
\end{center}

Just after that, an intermediary memory state, $h'_t$, is calculated according to this equation

\begin{center}
\begin{tabular}{r c l }
 $h'_t$ & $=$ &  \texttt{tanh}$(W . [r_t \times h_{t-1},x_t]) $\\  
\end{tabular}
\end{center}

The pointwise multiplication $r_t \times h_{t-1}$ allows the reset gate to decide, thanks to the sigmoid function in the definition of $r_t$, how much of the past hidden output $h_{t-1}$ to reset and how much to keep. The \texttt{tanh} function squeezes the intermediary hidden state between $-1$ and $1$.

The update gate $z_t$ is identical to the reset gate but differs in the weights. It also takes in entry the previous hidden output $h_{t-1}$ and the current input $x_t$, concatenates them  multiplies them by weights, and applies a sigmoid function to squeeze the output between 0 and 1. This is given by the equation

\begin{center}
\begin{tabular}{r c l }
 $z_t$ & $=$ & $\sigma(W_z . [h_{t-1},x_t] ) $ \\ 
\end{tabular}
\end{center}

Once done, the GRU cell updates the current  memory content according to this equation

\begin{center}
\begin{tabular}{r c l }
$h_t$ & $=$& $(1-z_t) \times  h_{t-1}+ z_t\times h'(t)$  
\end{tabular}
\end{center}

Note that if the update gate delivers a $z_t$  close to $0$, the current  memory content will be mostly influenced by the past value of $h_{t-1}$ and if it is close to $1$, it will be mostly influenced by the updated value $h'_{t}$.

In general, LSTM and GRU have comparable performance in prediction time and there is no universal way to prefer one over the other. However, GRUs are a little faster to train and need fewer data to generalize. Both LSTMs and GRUs tend to deprecate simple RNNs.


\subsubsection{Convolutional Neural Networks}  


Convolutional Neural Networks (CNNs) refer to a subcategory of neural networks. CNNs are mainly designed to process images but can also be used in other problems. Their architecture is more specific, though. It is composed of two main blocks. The first block is the particularity of this type of neural network, and functions as a feature extractor. To do this, it carries out a template-matching by applying convolution filtering operations. The first layer filters the input data with several convolution kernels and returns feature maps which are first normalized using an activation function and then resized. This process is repeated several times. This is to say,  a filtration of the feature maps obtained with new kernels is performed which gives birth to new feature maps to normalize and resize, which are filtered again, etc. Finally, the values of the last feature maps are concatenated in a vector. This vector is the output of the first block and the input of the second. The second block is not specific to a CNN. It is at the end of all the neural networks used for classification. The input vector values are transformed---using several linear combinations and activation functions--- to return a new vector as an output. This last vector contains as many elements as there are classes. The element $i$ generally represents the probability that the input belongs to the class $i$. Each element is therefore between $0$ and $1$ and the sum equals $1$. These probabilities are calculated by the last layer of this block---and therefore of the network---using a sigmoid function in case of two classes, or a Softmax function in case of multiple classes. As with ordinary neural networks, layer parameters are determined by the descent of the gradient and usually cross-entropy is minimized during the training phase. 

There are four types of layers \cite{CNNLayers} in a convolutional neural network: the convolution layer, the pooling layer, the ReLU correction layer and the fully-connected layer.

\begin{itemize}

\item The convolution layer (CONV): this layer is the key component of convolutional neural networks and is always at least the first layer. Its purpose is to identify the presence of a set of features in the data received as input (often images). This is achieved by convolution filtering. The principle is to slip a window representing the feature onto the data matrix and to calculate the convolution product of the feature and each portion by the scanned matrix. A feature is then seen as a filter. The convolution layer therefore receives several matrices as inputs, and calculates the convolution of each of them with each filter. The filters correspond exactly to the wanted features in the input matrix. For each pair (matrix, filter) we obtain an activation map, or feature map, which indicates where the features are located in the image. The higher the value, the more the corresponding location in the image looks like the feature;

\item The pooling layer (POOL): This type of layer is often placed between two convolution layers. It receives several feature maps as input, and applies to each of them the pooling operation which consists in reducing the size of the input matrices, while preserving their important characteristics. To do this, the input matrix is divided into regular cells, then the maximum value is kept within each cell. In practice, small square cells are often used to avoid losing too much information. The same number of feature maps is presented at the output as at the input, but they are much smaller. The pooling layer reduces the number of parameters and calculations in the network. This improves the efficiency of the network and reduces overfitting; and

\item The Rectified Linear Unit (ReLU) correction layer: this correction layer replaces all negative values received at the inputs with zeros using the ReLU activation function defined as follows.

$$\texttt{ReLU} (x) = \texttt{max}(0,x)$$ 

This results in neural networks several times faster without significantly impacting the generalization of the precision. This also increases non-linearity in the entire network. 

\item The fully-connected layer (FC): This is the last layer for every neural network whether it is convolutional or not. This type of layer receives an input vector and generates a new output vector. To do this, it applies a linear combination and possibly an activation function to the values received as inputs.

\end{itemize}




            






\newpage

\section{Advanced techniques}





%
\subsection{Ensemble Learning}        \label{secEnsembleLearnining}

\begin{figure}[h]
\centering
\includegraphics[scale=0.4,frame]{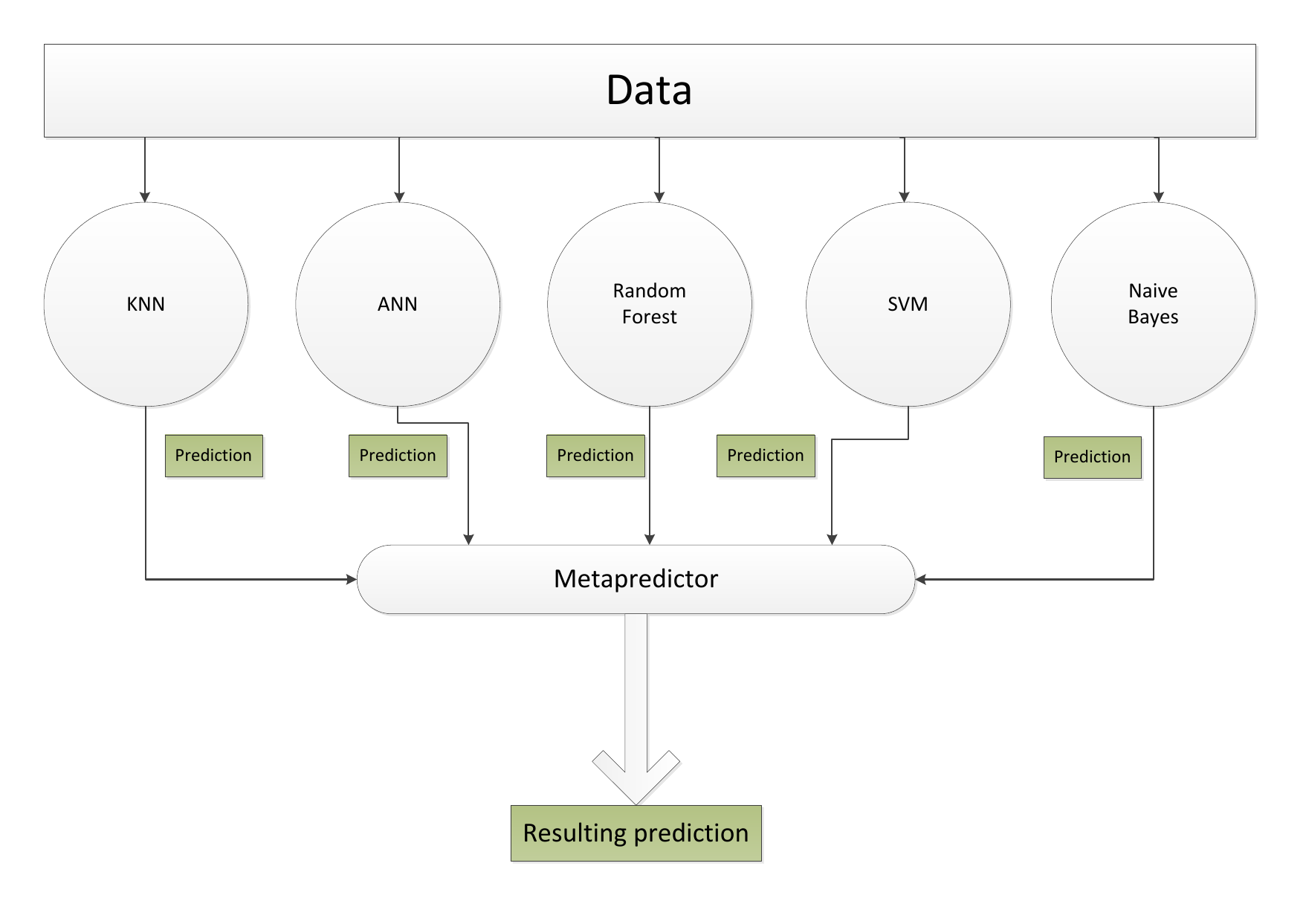}

\caption{Ensemble Learning process}
\label{EnsembleLearning2}
\end{figure}

\begin{figure}[h]
\centering

\includegraphics[scale=0.4,frame]{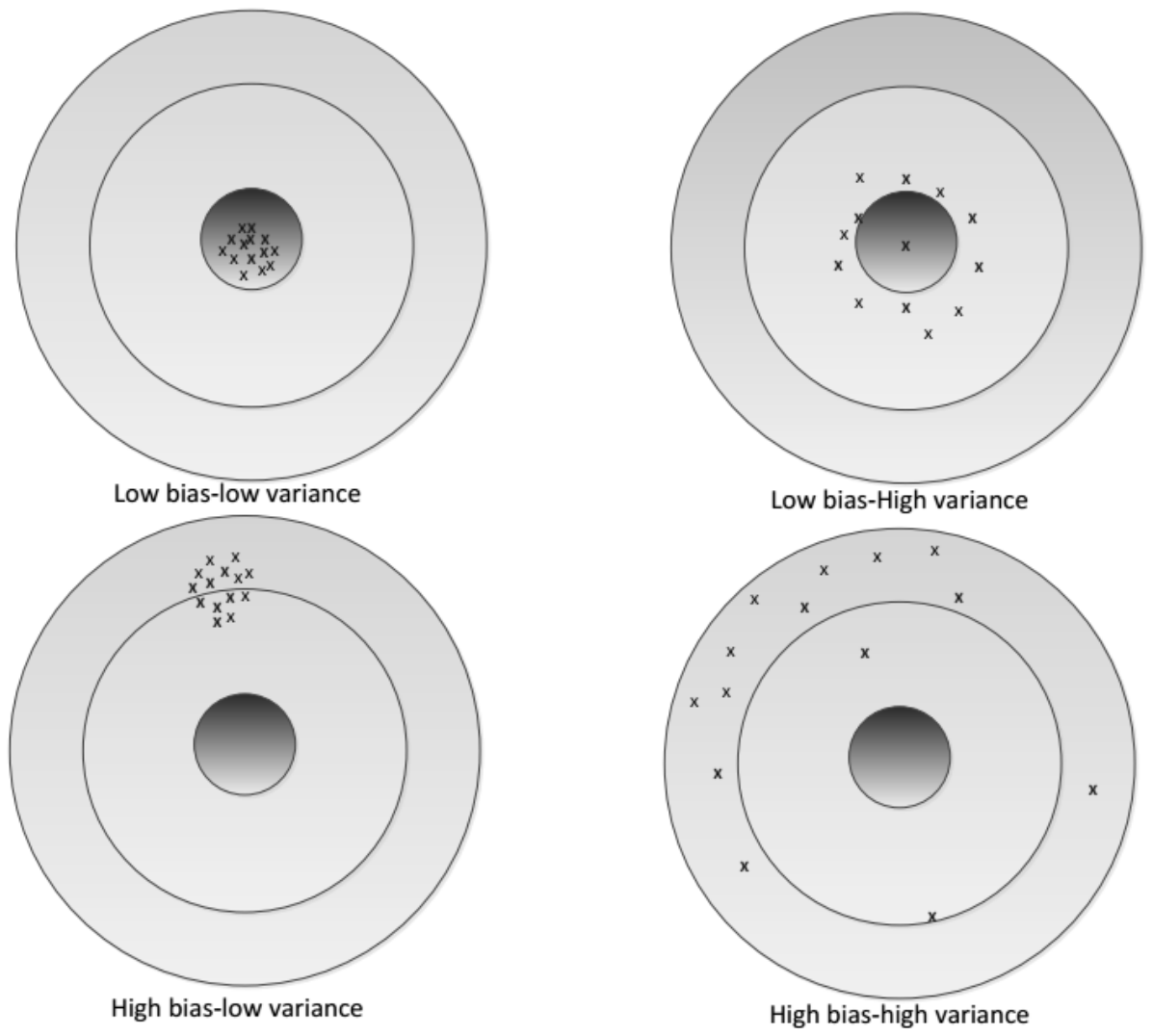}

\caption{Bias and variance}
\label{EnsembleLearning3}
\end{figure}

Ensemble Learning (EL) is the process of combining several decisions provided by several predictors to obtain the \textit{best} prediction (Figure \ref{EnsembleLearning2}).  
It basically tries to answer the question  stated by Kearns and Valiant\cite{Boosting}: "Can a set of weak learners create a single strong learner?".
EL is interested in determining the most strategic and efficient way to build a good metapredictor from less performing predictors. The main goal of EL is to minimize  the bias error and the variance of the separate models (Figure \ref{EnsembleLearning3}). The bias error is the average of the difference between the predicted result and the actual result. A high bias error implies that the model is under-performing and missing out on important patterns. On the other hand,  variance  quantifies how consistent the predictions are between each other over different training samples but does not care if the model is accurate or not.  Models that demonstrate high bias and low variance underfit the truth target. All the same, models that demonstrate low bias and high variance overfit the truth target. A trade-off between bias and variance  should be so established.


There are several EL techniques. Hereafter, we present the most commonly used ones. 

\subsubsection{Simple Ensemble Learning techniques}        
These techniques are straightforward and the decision could be one of the following:

\begin{enumerate}
\item The mode of the results, which is the most frequently occurring outcome of the used predictors;

\item The average of the results given by different models, if applicable;

\item The weighted average of the results  given by different models, if applicable. In this decision procedure,  a weight is assigned to each model and the final decision will be the weighted average. The weight could be the accuracy of the model observed on similar data.

\end{enumerate}

Although they are simple, these techniques happen to be very powerful in practice for many problems.


\subsubsection{Advanced Ensemble Learning techniques}        

\begin{enumerate}

\item Bootstrap AGGregating (BAGGing): in this technique, multiple random samples are first generated from the training dataset   with replacement, meaning that an element may crop up multiple times in the same sample. These samples are called bootstrapped samples. Then, an independent model is built for every sample using the same algorithm. Finally,  a combiner algorithm is used to aggregate over the decisions of these models (also called independent homogeneous weak learners) and a final decision is made, usually using average or plurality voting. Bagging hence allows multiple similar models with high variance to water down their variance;

\item Boosting: the fundamental difference between BAGGing and Boosting is that with the first technique weak homogeneous learners are trained in parallel whereas with Boosting they are trained in sequence in a way that \textit{each model learns from the mistakes of its predecessor}. That is, after each training step, misclassified data increases its weights so that the subsequent learner focuses on them during its training. Boosting also assigns a weight to each model in a way that a learner with a better classification result will be assigned a higher weight. In some Boosting techniques, a model with results judged very weak could be discarded.  By building multiple incremental models, Boosting usually leads to a decrease in the bias while keeping variance low; and


\item Stacking: Stacking is quite different from Boosting or BAGGing. Its main idea is to tackle a learning problem with different types of models that are able to learn some aspects of the problem, but not the entire problem. It uses a meta-model \textit{stacked} on top of a number of different intermediate learners. These  intermediate learners provide the meta-model with intermediate predictions, usually in the form of probabilities, which make up the its inputs.  According to their results, the meta-model will discern where each model performs well and where it performs poorly. In practice, the meta-model is a logistic regression. 

\end{enumerate}

It is worth mentioning that there is a practical package in Python called ML-Ensemble\cite{MLEnsemble} which simplifies the process of creating ensembles. This package is compatible with both Scikit-learn and Keras and so allows building ensembles for both deep and classical Machine Learning models.

\newpage

\section{Theory of evidence of Dempster-Shafer}        


Dempster-Shafer theory (DST) was first introduced by Arthur P. Dempster  \cite{Dempster2008},  then extended by Glenn Shafer \cite{shafer1976mathematical, SHAFER1990323, Yager8419294, Shekhar2019}. It has been proposed to formally model and validate the uncertainty associated with statistical inferences. This theory makes it possible to combine evidence stemming from different sources to reach a certain degree of belief on a given question. DST is also known as the theory of evidence. It is also perceived as a generalization of the Bayesian theory of subjective probability \cite{Smith2001}.  It has a clear connection with several  well-known theories such as probability theory \cite{jaynes2003}, possibility theory \cite{Dubois2006PTS16479671648302} and imprecise probability theory \cite{impreciseprobabilities}. All the same, it is fundamentally different from them.

\subsection{Formalism}     

Let $\Omega$ be a universe (i.e. the set of all entities that we want to consider) and $\mathcal{P}(\Omega)$ the set of all the subsets of $\Omega$ including the empty set $\emptyset$ (also denoted by $2^X$).

\paragraph{Mass function}     

The mass function, denoted by $m:$  $\mathcal{P}(\Omega)$    $\rightarrow$ $[0,1$], is a function such that $m(\emptyset) = 0$ and $\displaystyle\sum_{A\in \mathcal{P}(\Omega)}^{} m(A)= 1$. It expresses the proportion of available evidence supporting the proposition that an event $X$ is entirely $A$, where $A$ is a set of states. It is important to emphasize that the mass of $A$ does not consider the subsets of $A$, which have their own masses.

\paragraph{Belief function}    

The belief function, denoted by  $\texttt{bel}:$  $\mathcal{P}(\Omega)$    $\rightarrow$ $[0,1$], attributes to a set $A$ the sum of masses of all its subsets. Meaning, $\texttt{bel}(A)= \displaystyle\sum_{B\in \mathcal{P}(\Omega)|B\subseteq A}^{} m(B)$. This function expresses the proportion of available evidence supporting the observation that an event $X$ is \textit{something in} $A$.

\paragraph{Plausibility function}  

The plausibility function, denoted by  $\texttt{pl}:$  $\mathcal{P}(\Omega)$    $\rightarrow$ $[0,1$], attributes to a set $A$ the sum of masses of all the subsets that intersect $A$. Meaning, $\texttt{pl}(A)= \displaystyle\sum_{B\in \mathcal{P}(\Omega)|B\cap A\not = \emptyset}^{} m(B)$. This function expresses the fact that \textit{there is nothing that leads to conclude} that an event $X$ is not $A$.\\ $ $\\

 It is not hard to notice that: $$\texttt{pl}(A)= 1- \texttt{bel}(1- \overline{A})$$ and that under the Bayes conditions: $$\texttt{pl} = \texttt{bel}$$ It is  also easy to notice that: $$m(A) \leq \texttt{bel}(A) \leq  \texttt{pl}(A)$$

Table \ref{DST1} shows an example of calculation of mass, belief and plausibility in Dempster-Shafer theory.

\begin{table}[]
\centering

\begin{tabular}{|C{3.5cm}|C{2cm}|C{2cm}|C{3cm}|}
\hline
\ZZ \textbf{Hypothesis (attack type)} &   \textbf{Mass}  & \textbf{Belief}  & \textbf{Plausibility} \\\hline

\ZZ Empty set& $0$  & $0$  & $0$ \\\hline
\ZZ Denial-of-service (DDoS)&   $0.35$  & 0.35  & 0.67 \\\hline
\ZZ Man-in-the-middle (MitM) &   $0.15$    & 0.15  & 0.45 \\\hline
\ZZ SQL injection (SQL) &    $0.10$  & 0.10  & 0.38 \\\hline
\ZZ DDoS or MitM &   $0.12$  & 0.62  & 0.90 \\\hline
\ZZ DDoS or SQL   &   $0.10$  & 0.55  & 0.85 \\\hline
\ZZ MitM or SQL&   $0.08$  & 0.33  & 0.65 \\\hline
\ZZ DDoS or MitM or SQL   &    $0.1$  & 1  & 1 \\\hline

\end{tabular}

\caption{Example of mass, belief, and plausibility in Dempster-Shafer theory}
\label{DST1}
\end{table}

\paragraph{Joining masses}     

The need for joining masses arises when we have more than one source providing mass values (more than one predictor). Joining two masses consists in combining them in order to come up with a single overall value for the same set of states. This is given by the following rules known as Dempster's Rules of Combination (DRC)\cite{SHAFER201626} (given for two masses $m_1$ and $m_2$ stemming from two different sources)

$$
\text{DRC } :: \left\{
    \begin{array}{lll}
        m_{\texttt{joint}} (\emptyset)=  0\\ \\
        m_{\texttt{joint}} (A)=\displaystyle\frac{1}{1-k} \displaystyle\sum_{B \cap C = A\not = \emptyset}^{}m_1(B)m_2(C)\\
       \text{where } k = \displaystyle\sum_{B \cap C = \emptyset}^{} m_1(B)m_2(C), k\not = 1 ~~~~~~~~~~& & \\
    \end{array}
\right.
$$

$k$ is a factor that measures the conflict between two masses and $1-k$ is the normalization factor that allows one to ignore the conflict between the two masses involved.

\subsection{Usage in data classification}     

Although there are no fixed rules that define how to use DST in classification, most systems that do so follow the schema defined in Figure \ref{DST1}. To remain practical, we will present two ways, taken from the work of Chen et al. \cite{ChenWAR14}, of using this theory. Some irrelevant details are omitted for the sake of clarity.\\

\paragraph{First example}    

Authors use the Wisconsin Breast Cancer Diagnosis (WBCD) dataset   \cite{WBCD} which contains nine normalized integer attributes ranging from $1$ to $10$ and two classes: \texttt{normal} and \texttt{abnormal}. Smaller values tend to describe normality while larger values tend to describe abnormality. The dataset   is then randomly divided into $10$ subsets having almost equal sizes. For each run, one of the ten subsets is used as test data and the nine remaining subsets to train the model.  During the training process, for each run and for each attribute,  a modified median threshold $t$ is established. Then, for each attribute, the training data are ordered from small to large. Since the distribution of  the WBCD dataset   is a normal one, the mass functions, for each attribute, are modeled with a sigmoid function as follows:

$$
 \left\{
    \begin{array}{lll}
        m (\texttt{normal})= \displaystyle\frac{1}{1+e^{v - t}}\\ \\
        m (\texttt{abnormal})= 1 - m (\texttt{normal})\\
    \end{array}
\right.
$$

where $v$ is the value of the attribute. Then, DRC is used to calculate the overall mass values. Finally, for a given item, if the normal mass value is superior to the abnormal mass value, then it is classified normal. Otherwise, it is classified abnormal.

\begin{figure}[h]
\centering
\includegraphics[scale=0.4,frame]{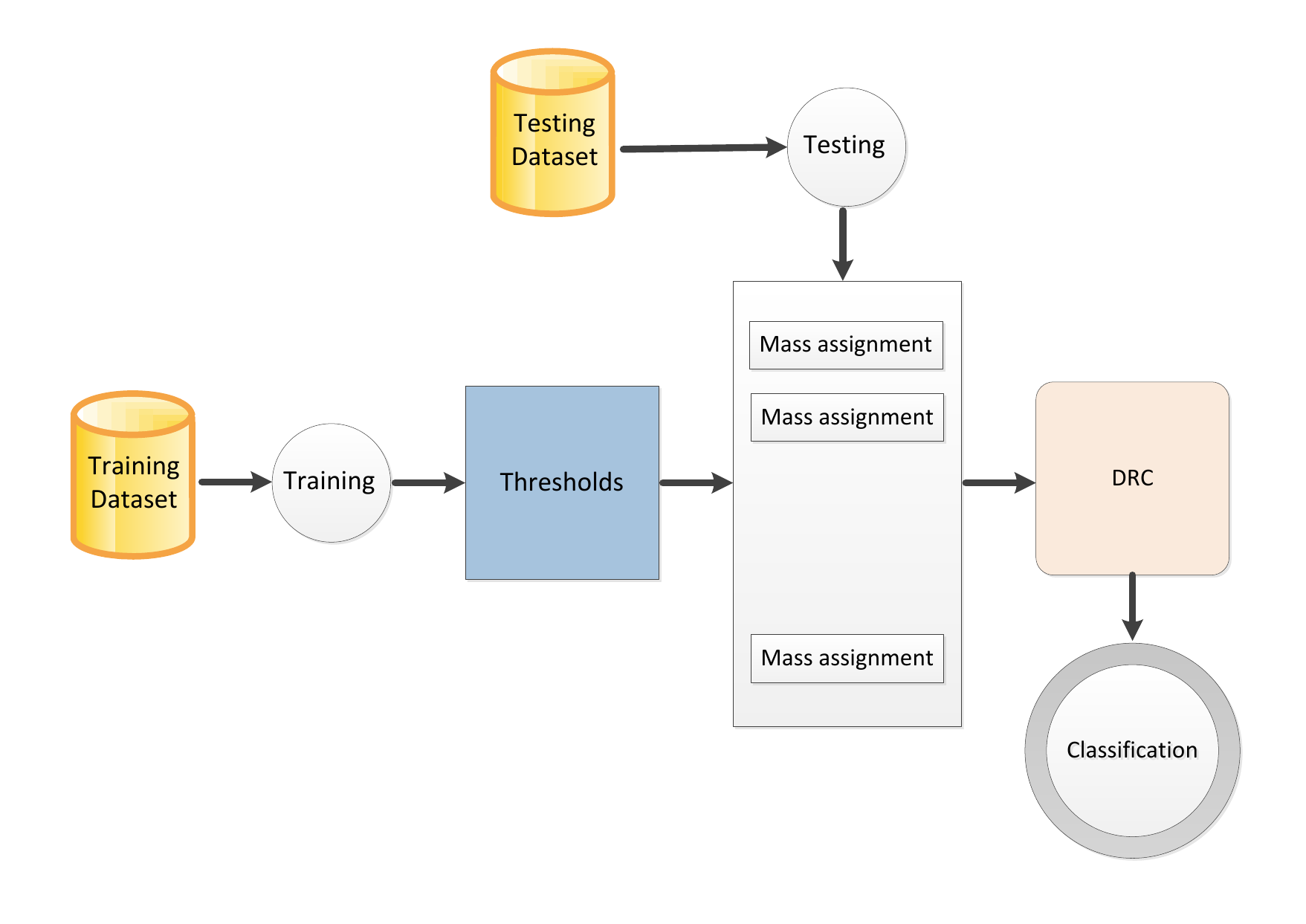}

\caption{DST exploitation scheme}
\label{DST1}
\end{figure}

\paragraph{Second example}     

In this example, the Iris dataset   \cite{IRIS} is used. The dataset   has four real positive attributes and three classes. The data are divided exactly the same way as in the previous example. During the training of one model, for each attribute, a maximum and a minimum value for each class are established. When an item can be classified into two classes (overlap between intervals), say \texttt{class1} and \texttt{class2}, then it is classified $\{\texttt{class1}, \texttt{class2}\}$ with a certain mass value. The same in case of three classes. The mass values are assigned as follows. When an item belongs to a single class, say \texttt{class1}, a value of $0.9$ is given to $m(\texttt{class1})$ and a value of $0.1$ is given to the uncertainty  $m( \texttt{any})$. Whereas when an item belongs to two classes \texttt{class1} and \texttt{class2}, a value of $0.9$ is given to $m(\{\texttt{class1, class2}\})$ and a value of $0.1$ is given to  $m( \texttt{any})$. Finally, when an item belongs to the three classes,  a value of $1$ is given to  $m( \texttt{any})$. After that, DRC is used to generate an overall hypothesis for the four attributes. The belief function is then used, and for each item, it is classified into the hypothesis with the highest belief value. If the hypothesis represents more than one class, a further step is performed. This step consists in calculating the Feature Selection Value (FSV), for each attribute for each class in the set of classes of the hypothesis. Then, the attribute $a$ having the smallest FSV is chosen. Then, for each of the $n$ candidate classes the distance $d$ between the attribute value $v_a$ and the mean value $\overline{v}_{a}$ of that attribute over the dataset  is calculated. The selected class is the one having the smallest distance. The  \texttt{FSV} and the distances $d_i$ are calculated as follows:

$$\texttt{FSV} =  \frac                 {\displaystyle \prod_{i =1}^{n}  \sigma (\texttt{class}_i)}                 { \displaystyle\sigma \bigcup_{i=1}^{n} \texttt{class}_i}$$

$$ d_i = |v_{a_i} - \overline{v}_{a_i}|, i \in \{1, ..., n\}$$

$\sigma \mbox{ is the standard deviation.}$\\ $ $\\

More elaborate examples can be found in \cite{ChenWAR14, WangLZ19a, AnHFZ19, ZhengLZLW19, ZengZWFW1918, GaoLY19}.

DS theory, in learning problems, is reputed to cope well with missing data. It is also known to enhance model accuracy with less overfitting. However, problem modelling using this theory remains data-dependent in terms of complexity. \\

\newpage

\section{Other related techniques}        

\vspace{0.5cm}

\subsection{Dropout}        

Dropout \cite{Srivastava2014} is perceived as a regularization technique to combat overfitting. Its principle is to randomly ignore a certain percentage of the nodes of a layer at each iteration. The final outgoing weights of each neuron are weighted according to its retention probability. This avoids the over-specialization of a neuron and therefore rote learning.

\subsection{Dropconnect}        

Dropconnect \cite {Wan2013} is an alternative to dropout consisting in inhibiting a connection and this always in a random way. In dropout, a number of randomly selected activated nodes is disabled in each layer whereas in dropconnect a randomly selected subset of weights in the network is temporarily zeroed (at each iteration). The results are similar to dropout in terms of speed and ability to generalize learning.

\subsection{Early stopping}        

Early stopping \cite{Yao2007} is a form of regularization used to combat overfitting. This technique allows the model parameters to be updated to better adapt to the training data for each iteration. To some point, this enhances the model's performance on data out of the training set. However, beyond this point, the improvement plays against the error generalization. Early stopping rules give indications of the number of iterations that can be performed before the model starts to overfit \cite{Prechelt2012}.

\subsection{Adversarial training}

Adversarial training tries to fool models using malicious input (adversarial examples). Such a technique can be useful, for example, to show how a malicious opponent can surreptitiously manipulate input data to exploit specific vulnerabilities in learning algorithms and jeopardize the security of the Machine Learning system. The adversarial examples are intentionally constructed using an optimization procedure \cite{HuangPGDA17}. They help to regularize models and make them more robust.


\subsection{Data Augmentation}        


The easiest way to reduce overfitting is to increase the size of the training dataset. However, doing so may turn out to be too expensive. For that, some research studies have emerged trying to reduce this cost by proposing augmentation algorithms and policies. For example, D. Cubuk et al. in \cite{DBLP:Cubuk} propose a  procedure called AutoAugment to automatically search for improved data augmentation policies. Population-Based Augmentation (PBA) policy \cite{PBA2019} has also been proposed and  shown to have better performance than AutoAugment with one thousand times less computations in some cases. All the same, this result should be taken with caution pending confirmation by further research.

\newpage

\section{Practical Aspects}                


\vspace{0.5cm}

\subsection{Python}

Python is a programming language for general use but very suitable for Machine Learning thanks to multiple dedicated packages and libraries. Among them we mention the following.

\subsubsection{Scikit-Learn}

Scikit-Learn\cite{scikit-learn} is a Python library for Machine Learning that encompasses functions for estimating algorithms such as random forests, logistic regressions, classification algorithms, and support vector machines. It offers great advantages, such as:

\begin {itemize}
    \item It is an open source project;
    \item Its code is fast, some parts are implemented in Cython;
    \item It has excellent documentation and a large and active community;
    \item It is very well integrated with the Pandas and Seaborn Libraries;
    \item It has a uniform API between all algorithms, making it easy to switch from one to the other;
\end {itemize}

For a new programmer in the Machine Learning domain, it is recommended that they have good knowledge of the following packages:

\begin {itemize}
    \item Scipy:  Scipy \cite{SciPy} is an open-source ecosystem for mathematics, science, and engineering.  It offers a large number of numerical routines for integration, interpolation, optimization, linear algebra, and statistics;
    \item Pandas: Pandas \cite{pythonpandas} is also an open-source library offering high-performance and easy-to-use data structures, as well as data analysis tools;
    \item Numpy: Numpy \cite{numpy}  is an essential package for scientific computing with Python. It provides powerful capabilities for N-dimensional array management, linear algebra calculus, Fourier transformation handling, random number utilization, etc.;
    \item Matplotlib: Matplotlib \cite{Matplotlib} is a fundamental Python library  to generate 2D graphics such as lines, curves, histograms, bar charts, error charts, scatter plots, etc.
    
\end {itemize}

\subsubsection{Tensorflow}

Tensorflow\cite{tensorflow2015-whitepaper} is an open source Machine Learning library, created by Google, for developing and running Machine Learning and Deep Learning applications. It can be conceived as a programming system in which calculations are represented in the form of graphs. The nodes represent mathematical operations, and the arrows represent multidimensional data exchanged between them. Although Tensorflow is a Python library, the mathematical operations themselves are not performed in Python. They are written in high-performance C++. Thanks to abstraction, this framework facilitates the implementation of algorithms and allows the developer to focus on the general logic of an application. In addition, the eager execution mode allows the evaluation and modification of all graph operations separately rather than having to build the entire graph as a single opaque object and evaluate everything at once. Besides, the TensorBoard visualization suite allows the inspection of graphics' internal workings through a web-based interactive dashboard. Among its disadvantages is the fact that the same model running on two different systems will sometimes vary even if the data with which it has been fed are exactly the same. It also requires knowledge of advanced calculus and linear algebra as well as an important understanding of the Machine Learning complexity.

\subsubsection{Theano}

Theano \cite{Theano} is a numerical computation  and Deep Learning Python library developed by Mila, a research team from McGill University and the Université de Montréal. It support tensors and algebra operations and allows  parallel execution. However, it may output unclear error messages. In addition, compiling a model in Theano may take a long time compared to other libraries. Theano, nowadays, tends to be masked by Tensorflow which is gaining a lot of ground to the detriment of Theano.

\subsubsection{CNTK}

Microsoft Cognitive Toolkit (CNTK) \cite{CNTK} is a Python framework for Machine Learning and Deep Learning developed by Microsoft Research. It provides an intuitive way to build models and offers easy interfaces. It is an alternative to Tensorflow.  

\subsubsection{Torch}

Torch \cite{Torch} is a software platform for scientific computing that primarily uses GPUs and allows one to work with multiple Machine Learning algorithms. Its ease of use and efficiency are due to the LuaJIT scripting language and the underlying implementation of C/CUDA. It is worth mentioning that Torch is not the same project as PyTorch, despite the similar name.

\subsubsection{PyTorch}

PyTorch \cite{PyTorch} is an open-source Python project similar to Torch that provides several Machine Learning algorithms. It also offers a  library called LibTorch to implement extensions to PyTorch and write Machine Learning applications in pure C++. Models written with PyTorch can be converted and used in C++ with TorchScript.

\subsubsection{Keras}

Keras \cite{chollet2015} is a high level, user-friendly, modular, and extensible API that facilitates the development of Deep Learning applications and can run on Tensorflow, Theano and CNTK on both CPUs and GPUs. Keras is part of the Tensorflow core, making it Tensorflow's preferred API.





\subsection{R}    


R\cite{MachineLearningWithR} is one of the main languages of statistics and data science. Many statistical and Machine Learning methods are implemented in R, organized in specialized packages like rpart, mice, party, caret, randomForest, nnet, e1071,  kernLab, igraph, glmnet, ROCR, gbm, arules, tree, klaR, ipred, lars, CORElearn, and earth.  R  comes with a large number of experimental datasets and has  an interface to Keras via the libraries kerasR, Keras, and keras. R is a serious rival to Python and is used by a growing number of data engineers.

              
\subsection{Matlab}  

Matlab \cite{MachineLearningWithMatlab} implements several Machine Learning  algorithms such as  logistic regression, classification trees, support vector machines, ensemble methods, and Deep Learning. It also uses model refinement and reduction techniques to create accurate models. Nevertheless, Python and R seem to be easier and faster overall, but from a graphical and visual point of view, Matlab seems to be more convenient.


\subsection{Java}       

There are several Java environments that implement Machine Learning algorithms. Among them, we mention the following.

\begin{itemize}


\item Weka \cite{weka}: Weka stands for Waikato Environment for Knowledge Analysis.  It is open source software  developed by the University of Waikato, New Zealand. It consists of a collection of Machine Learning and Deep Learning algorithms and encompasses hands-on tools for data preparation, classification, regression, clustering, and visualization;

\item KNIME \cite{KNIME}: KNIME stands for KoNstanz Information Miner.  It is open source software  developed by Konstanz University, Germany. It is an integrated tool that implements a wide range of Machine Learning algorithms as well as genetic algorithms and may be used along with Java tools like Spark. The KNIME Deep Learning extensions allow users to read, create, edit, train, and execute deep neural networks and permit the integration with Tensorflow using both CPUs and GPUs;

\item RapidMiner \cite{RapidMiner}: RapidMinEr is developed at Technical University of Dortmund, Germany. It is a free open source software dedicated to data engineering. It contains many tools to process data: reading different input formats, data preparation and cleaning, statistics, Machine Learning and Deep Learning algorithms, performance evaluation, and various visualizations. It is not easy to handle at first sight, but with a little practice, it allows to quickly set up a complete data processing chain, from data entry to their classification;

\item MALLET \cite{MALLET}: MALLET stands for  MAchine Learning for LanguagE Toolkit. It is basically a Java-based package for Machine Learning applications. It also encompasses routines and tools for text processing,  sequence tagging, natural language processing, information extraction, and numerical optimization;

\item DL4J \cite{DL4J}: DL4J stands for Deep Learning For Java. It is the first professional open source commercial Deep Learning library written for Java and Scala developers. Integrated with Hadoop and Spark, DL4J is designed for use in professional environments with GPU and distributed processors;

\item Java-ML \cite{JavaML}: Java-ML is a Java API with a collection of automatic learning algorithms implemented in Java. It does not provide any standard interface for algorithms;

\item JSAT \cite{JSAT}: JSAT is a library to quickly start working with automatic learning problems. It is available for use under GPL 3. JSAT has no external dependencies, and is pure Java;

\item MLlib \cite{MLlib}: MLlib is a scalable Spark Apache learning library. It contains a long list of Machine Learning algorithms and  several underlying statistical, optimization, and linear algebra primitives.

\end{itemize}









\newpage
\section{Conclusion}               

The advancements in cybersecurity and digital forensics brought about a change through machine learning and deep learning technologies that offer ways to tackle intricate issues like identifying threats and anomalies in data analysis and studying digital evidence, with improved methods. These approaches have shown promise in improving precision and effectiveness while also addressing issues such as resilience and the protection of data. Our future focus will be on transfer learning \cite{trans1,trans2,trans3} which makes use of existing models in specialized tasks and investigating the capabilities of Graph Neural Networks (commonly referred to as GNNs) particularly used in depicting intricate cyber landscapes and digital connections. We will also be delving into Explainable AI (known as XAI) \cite{Manaiex1,Manaiex2,Manaiex3,Manaiex4,Manaiex5,Manaiex6} which enhances model transparency and interpretability.

\bibliographystyle{unsrt}
\bibliography{biblio.bib}

\appendix

\chapter{List of the most important conferences}

In Table \ref{Conferences}, we give a list of the most famous international conferences focusing on security and cyber security as well as their rank. We rely on the COmputing Research and Education  (CORE) association of Australasia \cite{CORE} which is one of the most reliable conference and journal ranking institutions in the academic world. According to CORE, conferences are assigned to one of the following categories:

\begin{itemize}
    \item A* : flagship conference, a leading venue;
    \item A  : excellent conference, and highly respected;
    \item B  : good conference, and well regarded;
    \item C  : conference that meets minimum standards.
\end{itemize}
    
The provided list may not be exhaustive and other important conferences could be added.

\begin{table}
\begin{tabular}{|C{7cm}|C{3cm}|C{2cm}|}
\hline
\Z \textbf{Conference name} &  \textbf{Initialism}  &  \textbf{Rank}  \\
\hline
\N ACM Conference on Computer and Communications Security & CCS & A*  \\
\hline
\N IEEE Symposium on Security and Privacy & S\&P  & A*  \\
\hline

\N Usenix Security Symposium & USENIX-Security  & A*  \\
\hline

\N IEEE Computer Security Foundations Symposium &   CSF (was CSFW) & A  \\
\hline

\N European Symposium On Research In Computer Security &   ESORICS & A  \\
\hline

\N IEEE/IFIP International Symposium on Trusted Computing and Communications&   TrustCom & A  \\
\hline

\N Annual Computer Security Applications Conference &  ACSAC & A  \\
\hline

\N IEEE Conference on Systems, Man and Cyber netics&   SMC & B  \\
\hline

\N IFIP Information Security \& Privacy Conference&    IFIP SEC & B  \\
\hline

\N International Conference on Availability, Reliability and Security&   ARES & B  \\
\hline

\N ACM Symposium on Information, Computer and Communications Security&   AsiaCCS & B  \\
\hline

\N Computational Intelligence in Security for Information Systems&   CISIS & B  \\
\hline

\N Information Security Practice and Experience Conference&   ISPEC & B  \\
\hline

\N IEEE Symposium on Computational Intelligence in Cyber  Security&    IEEE CICS & C  \\
\hline

\N International Workshop on Visualization for Cyber  Security&    VizSec & C  \\
\hline

\N International Conference on Computational Intelligence and Security&    CIS & C  \\
\hline

\N International Conference on Information Security and Cryptography&    SECRYPT & C  \\
\hline

\N International Conference on Risks and Security of Internet and Systems&    CRiSIS & C  \\
\hline

\N International Workshop on Critical Information Infrastructures Security&    CRITIS & C  \\
\hline

\end{tabular}

\caption{List of the most important conferences}
\label{Conferences}
\end{table}

\chapter{List of the most important journals}

In Table \ref{Journaux}, we give a list of some famous journals for security and Machine Learning applications and techniques. The journals are ranked by the Computing Research and Education Association of Australasia (CORE) \cite{CORE}. Other journals of high quality could also be added.

\begin{table}
\begin{tabular}{|C{7.5cm}|C{3cm}|C{2cm}|}
\hline
\Z  \textbf{Journal name} &  \textbf{ISSN}  &   \textbf{Rank}  \\
\hline

\N IEEE Transactions on Neural Networks and Learning Systems (was IEEE Transactions on Neural Networks) & 1045-9227 & A*  \\
\hline

\N ACM Transactions on Computer Systems & 0734-2071 & A*  \\
\hline

\N ACM Transactions on Software Engineering and Methodology & 1557-7392 & A*  \\
\hline

\N IEEE Transactions on Computers & 0018-9340 & A*  \\
\hline

\N IEEE Transactions on Evolutionary Computation & 1941-0026 & A*  \\
\hline

\N Machine Learning & 1573-0565 & A*  \\
\hline

\N ACM Transactions on Information and System Security & 1094-9224 & A  \\
\hline

\N IEEE Transactions on Information Forensics and Security & 1556-6013 & A  \\
\hline

\N IEEE Transactions on Dependable and Secure Computing & 1545-5971 & A  \\
\hline

\N IEEE Transactions on Knowledge and Data Engineering & 1558-2191 & A  \\
\hline

\N Journal of Machine Learning Research & 1532-4435 & A  \\
\hline

\N Computers and Security & 0167-4048 & B  \\
\hline

\N Journal of Computer Security & 1875-8924 & B  \\
\hline

\N IEEE Security and Privacy Magazine & 1540-7993 & B  \\
\hline

\N Information Security Technical Report & 1363-4127 & B  \\
\hline

\N IEEE Transactions on Systems, Man and Cybernetics, Part A: Systems and Humans & 1083-4427 & B  \\
\hline

\N IET Information Security &  1751-8709 & C  \\
\hline

\N Information Security Journal: a Global Perspective &  1939-3555 & C  \\
\hline

\N International Journal of Security and Networks &  1747-8405 & C  \\
\hline

\N International Journal of Information Security &  1615-5262 & C  \\
\hline

\N International Journal of Electronic Security and Digital Forensics &  1751-9128 & C  \\
\hline

\end{tabular}

\caption{List of the most important Journals}
\label{Journaux}
\end{table}

\chapter{List of the most significant books}

Hereafter, we give a list of the most relevant books related to the theme of our work in this document.

\begin{itemize}

\item Fundamentals of Machine Learning for Predictive Data Analytics: Algorithms,  \\ Worked Examples, and Case Studies, by John D. Kelleher (Author), Brian Mac Namee (Author), Aoife D'Arcy (Author).    ISBN: 978-0262029445. Publisher: The MIT Press (2015).

\item Support Vector Machines, by Ingo Steinwart (Author), Andreas Christmann (Author). ISBN: 978-0387772417. Publisher: Springer (2008).

\item Deep Learning with Python, by Francois Chollet (Author).  ISBN: 978-1617294433. Publisher: Manning Publications (2017).

\item Hands-On Machine Learning with Scikit-Learn and TensorFlow: Concepts, Tools, and Techniques to Build Intelligent Systems, by Aurélien Géron (Author). ISBN: 978-1491962299. Publisher: O'Reilly Media (2017).

\item Deep Learning, by Ian Goodfellow (Author), Yoshua Bengio (Author), \\ Aaron Courville (Author).  ISBN: 978-0262035613. Publisher: The MIT Press (2016).

\item Machine Learning: An Algorithmic Perspective, by Stephen Marsland (Author).  ISBN: 978-1466583283. Publisher: Chapman and Hall/CRC (2014). 

\item Machine Learning: A Bayesian and Optimization Perspective, by Sergios Theodoridis (Author).  ISBN: 978-0128015223. Publisher: Academic Press (2015).

\item Machine Learning: A Probabilistic Perspective, by Kevin P. Murphy (Author).  ISBN: 978-0262018029. Publisher: The MIT Press (2012).

\item The Elements of Statistical Learning: Data Mining, Inference, and Prediction, by Trevor Hastie (Author), Robert Tibshirani (Author), Jerome Friedman (Author). ISBN: 978-0387848570. Publisher: Springer (2017).

\item Deep Learning Applications for Cyber  Security, by Mamoun Alazab (Editor), \\ Ming Jian Tang (Editor).  ISBN: 978-3030130565. Publisher: Springer (2019).

\item Classic Works of the Dempster-Shafer Theory of Belief Functions, by Ronald R. Yager (Editor), Liping Liu (Editor). ISBN: 978-3642064784. Publisher: Springer (2008). 

\item Combining Classifiers using the Dempster Shafer Theory of Evidence, by Imran Naseem (Author). ISBN: 978-3639232240. Publisher: VDM Verlag Dr. Müller (2010). 

\item Understanding Machine Learning: From Theory to Algorithms, by \\ Shai Shalev-Shwartz (Author), Shai Ben-David (Author).  ISBN: 978-1107057135. Publisher: Cambridge University Press (2014).

\item Machine Learning: The Art and Science of Algorithms that Make Sense of Data, by Peter Flach (Author).  ISBN: 978-1107422223. Publisher: Cambridge University Press (2012).

\item Cyber security Essentials,  by Charles J. Brooks (Author), Christopher Grow (Author), Philip Craig (Author), Donald Short (Author).  ISBN: 978-1119362395. Publisher: Sybex (2018).

\end{itemize}

\chapter{List of the most relevant online tutorials}

On the Internet, there is a large number of resources that offer reliable information and tutorials on Machine Learning and Deep Learning. Among them, we choose the following resources:

\begin{itemize}[-]

\item Udemy: Udemy is a  marketplace for teaching and learning. It offers online courses on a wide spectrum of knowledge, in particular in the field of Reinforced Learning, Unsupervised and Supervised Learning, Python, Keras, AI, ML, DL, etc.;

\item Datacamp: Datacamp is one of the most well-known platforms for learning Data Science, ML, and DL online;

\item Luis Serrano tutorials: Luis Serrano is a Machine Learning Engineer at Google. He offers free video sequences on several aspects of the field in a simplified way. His channel could easily be found on Youtube;

\item MIT Deep Learning: it is a collection of MIT courses and lectures on Deep Learning, deep reinforcement learning, autonomous vehicles, and artificial intelligence; and

\item Tutorialspoint: Tutorialspoint is one of the biggest online tutorials libraries. It's free and generally aimed at beginners;

\end{itemize}

\chapter{List of the most used websites for datasets}

Kaggle offers a large number of public datasets in a large number of fields. These datasets are listed here \cite{kaggle2019}. Other public datasets are available at Data World \cite{dataworld2019}, Data Gov   \cite{datagov2019}, Socrata OpenData  \cite{SocrataOpenData2019}, UCI Machine Learning Repository \cite{UCI2019}, Amazon Web Services Open Data Registry \cite{Amazon2019}, Earthdata \cite{Earthdata2019}, and Reddit \cite{RDatasets} websites; and many others.

\chapter{SVM Theory}


\subsection*{Notations and definitions}

\subsubsection*{Norm of a vector}

The norm (or magnitude) of a vector $x=(x_1, x_2, ..., x_n)$ in a vector space of dimension $n$ is its length. It is denoted by: $$||x|| = \sqrt{x_1^2 + x_2^2+ ... + x_n^2}$$

\subsubsection*{Direction of a vector}

The direction of a vector $x=(x_1, x_2, ..., x_n)$, denoted by $w$, is the vector:
$$w=(\frac{x_1}{||x||}, \frac{x_2}{||x||}, ..., \frac{x_n}{||x||})$$

Geometrically, $w$ represents in which direction vector $x$ is oriented with no regard to its norm. Parallel vectors have the same direction.

\subsubsection*{Dot product}

The dot (or scalar) product of two vectors $x=(x_1, x_2, ..., x_n)$ and $y=(y_1, y_2, ..., y_n)$ is 
$$x.y = \displaystyle\sum_{i=1}^{n}  x_i.y_i$$

The dot product is a real number not a vector. Geometrically, it is the product of the Euclidean norms of the two vectors and the cosine of the angle between them. It represents \textit{how much push one vector is giving to the other}.

\subsection*{Linear separation}

\subsubsection*{Hyperplane}

In a two dimensional space, data might be separated by a line having the equation $y = a.x+b$ or simply by $a.x-y+b=0$. If we replace $x$ with $x_1$ and $y$ with $x_2$, the equation becomes 
$$ a.x_1 -x_2 +b = 0$$ 

Let $x = (x_1,x_2)$ and $w=(a,-1)$, we obtain:

$$w.x+b = 0$$

This equation can be generalized to a space of dimension $n$. In that case, it is called the equation of hyperplane.

\subsubsection*{Classification}

If we know the equation of the hyperplane that segregates data, the classification is given by the following  function for any vector $x_i$

$$
 h(x_i) = \left\{
    \begin{array}{ll}
       +1 & \mbox{if } w.x_i+b \geq 0 \\
       -1 & \mbox{else.}
    \end{array}
\right.
$$

That means that the points above the hyperplane are classified as class +1, and the points below the hyperplane are classified as class -1.  The problem is then how to find the best hyperplane. This will require  hyperplanes' metrics.

\paragraph*{Bad metric: A}

Let us have a dataset of $m$ rows defined as follows:

$$D= \{(x_i,y_i) | x_i \in \bm{\mathbb{R}}^n \wedge y_i \in \{-1, +1\}\}$$

Let us have a hyperplane of equation $w.x+b = 0$.

To determine the closest point point to the hyperplane, let us define the metric $A$ such that:

$$A = \displaystyle \min_{i\in\{1, ..., m\}}|w.x+b|$$

If we have $j$ hyperplanes, every hyperplane will have a value for $A$. The intuitive selection would be to choose the hyperplane corresponding to the largest $A$ value (the furthest-nearest point in order to maximize the margin) defined by:

$$H =  \displaystyle \max_{i\in\{1, ..., j\}}\{h_i|A_i\} $$

However, due to the absolute value in the definition of the metric $A$---that does not distinguish between a point located on one side of the hyperplane and another point on the other side, and then may lead to a bad classification, this metric could not be used as is.

\paragraph*{Better metric: B}

Now, instead of using $A$, we use the function $B$ defined as follows:

$$B = \displaystyle \min_{i\in\{1, ..., m\}} y_i(w.x+b)$$

Unlike $A$, $B$ uses the label of $y$ (i.e. $-1$ or $+1$) in its definition. Like that, the product $y_i(w.x+b)$ will always be positive for a good classification and always negative for a bad one. We  now may be tempted to select the hyperplane with the largest positive value of $B$ defined by:

$$H =  \displaystyle \max_{i\in\{1, ..., j\}}\{h_i|B_i\} $$

Unfortunately, $B$ is still not good as a metric as it suffers from the scale variant problem. That is, if we consider $w_1 = (1,2)$ and $w_2=(2,4)$, although they do represent the same hyperplane (collinear vectors), they can give different results when dot-multiplied by another vector.

\paragraph*{Best metric: C}

Now, instead of using $B$, we use the function $C$ defined as follows

$$C = \displaystyle \min_{i\in\{1, ..., m\}}y_i(\frac{w}{||w||}.x+\frac{b}{||w||})$$

In the definition of $C$, dividing by $||w||$ remedies the  scale variant problem. We finally  select the hyperplane with the largest positive value of $C$ defined by:


$$H =  \displaystyle \max_{i\in\{1, ..., j\}}\{h_i|C_i\} $$

\subsubsection*{Optimization}

We are now led to find the hyperplane $H$, fully defined by $w$ and $b$, that maximizes $C$ provided that all the points with the same class in the $m$ entries in the dataset have a distance from $H$ larger than or equal to $C$. This is expressed by the following optimization problem (s.t. means subject to):

$$
\begin{array}{ccc}
&\displaystyle \max_{w,b}  {{ ~C}} &\\
&\textrm{s.t.} y_i(\frac{w}{||w||}.x+\frac{b}{||w||}) \geq C, i \in \{1 ... m\} & \\
\end{array}
$$

Considering $C = \frac{B}{||w||}$, the problem becomes:

$$
\begin{array}{ccc}
&\displaystyle \max_{w,b}  {\frac{B}{||w||}} &\\
&\textrm{s.t.} y_i(w.x+b) \geq B, i \in \{1 ... m\} &\\
\end{array}
$$

Let us rescale $w$ such that $B=1$. The problem becomes:

$$
\begin{array}{ccc}
&\displaystyle \max_{w,b}  {\frac{1}{||w||}} &\\
&\textrm{s.t.} y_i(w.x+b) \geq 1, i \in \{1 ... m\} &\\
\end{array}
$$

Which can be expressed as:

$$
\begin{array}{ccc}
&\displaystyle \min_{w,b}  {||w||} &\\
&\textrm{s.t.} y_i(w.x+b) \geq 1, i \in \{1 ... m\} &\\
\end{array}
$$

This problem is equivalent to  the convex quadratic optimization problem expressed as follows:

\begin{equation}
\begin{array}{ccc}
&\displaystyle \min_{w,b}  {\frac{1}{2}||w||^2} &\\
&\textrm{s.t.} y_i(w.x+b) -1 \geq 0, i \in \{1 ... m\} &\\
\end{array}
\end{equation}

\paragraph*{Solving the SVM convex quadratic optimization problem} 

We give here a short overview on the solution of the SVM convex quadratic optimization problem\cite{ConvexOptimization}. A more detailed explanation can be found in \cite{B918972F, SVMmath}. 

The SVM convex quadratic optimization problem can be solved by the Lagrange method. Because it is quadratic, the surface is a paraboloid (convex) with a single minimum. This avoids the problem of local optimums encountered with neural networks. The Lagrange dual problem\cite{wiki:Duality} states that solving the convex problem: 
$$
\begin{array}{ccc}
&\displaystyle \min_{x \in D}  f(x)&\\
&\textrm{s.t.} g(x) \leq 0, i\in\{1,...,m\} &\\
\end{array}
$$

comes down to solving the problem:

$$
\begin{array}{ccc}
&\displaystyle \max_{\lambda} \min_{x}  ~ (f(x)+\displaystyle\sum_{i=1}^{m}\lambda_ig_i(x))&\\
&\textrm{s.t.} \lambda_i \geq 0, i\in\{1,...,m\} &\\
\end{array}
$$

The objective function $f(x)+\displaystyle\sum_{i=1}^{m}\lambda_ig_i(x)$ is called the Lagrangian function and denoted by $\mathcal{L}$, and $\lambda_i$ are called the Lagrange multipliers. In the case of the SVM convex optimization problem, we have:

$$
 \left\{
    \begin{array}{lll}
        f(w) & = &  \frac{1}{2}||w|| ^2\\
         g_i(w,b )& = & y_i .(w.x+b) -1, i\in\{1,...,m\} \\
         \mathcal{L}(w,b,\lambda)  & = & \frac{1}{2}||w||^2 - \displaystyle\sum_{i=1}^{m}\lambda_i  [y_i (w.x+b) -1] \\
    \end{array}
\right.
$$

The Lagrange dual problem for  the SVM convex optimization problem is then:
\begin{equation}
\begin{array}{ccc}
&\displaystyle \max_{\lambda} \min_{w,b}  ~ \frac{1}{2}||w||^2 - \displaystyle\sum_{i=1}^{m}\lambda_j  [y_i .(w.x+b) -1]&\\
&\textrm{s.t.} \lambda_i \geq 0, i\in\{1,...,m\} &\\
\end{array}
\end{equation}

There are some dedicated packages to solve convex optimization problems such as CVXOPT\cite{CVXOPT} and CVXPY\cite{CVXPY} in Python, CVX\cite{CVX} in Matlab and GUROBI\cite{GUROBI} in Java.

Another way to solve the SVM optimization problem is to solve its Wolfe dual form \cite{wiki:Duality} stated as follows:

\begin{equation}
\begin{array}{clc}
&\displaystyle \max_{\lambda,w,b}  ~ \frac{1}{2}w.w - \displaystyle\sum_{i=1}^{m}\lambda_i  [y_i (w.x_i+b) -1]&\\
&\textrm{s.t. } \mathop{\nabla} \frac{1}{2}w.w -  \displaystyle\sum_{i=1}^{m}\lambda_i \mathop{\nabla} [y_i (w.x_i+b) -1]=0&\\
&\textrm{and } ~ \lambda_i \geq 0, i\in\{1,...,m\} &\\
\end{array}
\end{equation}

where $\mathop{\nabla}$ is the gradient and $\lambda_i$ are the Lagrange multipliers. The Lagrangian function to maximize is: 


\begin{equation}
\begin{array}{ c c l}
 \mathcal{L}(w,b,\lambda) & = & \frac{1}{2}w.w - \displaystyle\sum_{i=1}^{m}\lambda_i  [y_i (w.x_i+b) -1]\\ 
& =  & \frac{1}{2}w.w - \displaystyle\sum_{i=1}^{m}\lambda_i  y_i w.x_i -   b \displaystyle\sum_{i=1}^{m}\lambda_i  y_i  +   \displaystyle\sum_{i=1}^{m}\lambda_i 
\end{array}
\label{eqwolf1}
\end{equation}

We calculate the gradient over $w$ and $b$, we get from the constraint:

\begin{equation}
\begin{array}{ c c l}
\mathop{\nabla}_w \mathcal{L}(w,b,\lambda) & = & w -  \displaystyle\sum_{i=1}^{m}\lambda_i y_i x_i = 0\\ 
\mathop{\nabla}_b \mathcal{L}(w,b,\lambda) & =  & - \displaystyle\sum_{i=1}^{m}\lambda_i  y_i = 0
\end{array}
\label{eqwolf2}
\end{equation}

So we get:

\begin{equation}
\begin{array}{ c c l}
w & = & \displaystyle\sum_{i=1}^{m}\lambda_i y_i x_i\\ 
 \displaystyle\sum_{i=1}^{m}\lambda_i  y_i & =  &  0
\end{array}
\label{eqwolf3}
\end{equation}

Considering the equation \ref{eqwolf3}, the equation \ref{eqwolf1} becomes:

\begin{equation}
\begin{array}{ c c l}
 \mathcal{L}(w,b,\lambda) & =  & \frac{1}{2}w.w - \displaystyle\sum_{i=1}^{m}\lambda_i  y_i w.x_i -   b \displaystyle\sum_{i=1}^{m}\lambda_i  y_i  +   \displaystyle\sum_{i=1}^{m}\lambda_i \\
 & =  & \frac{1}{2}w.w - \displaystyle\sum_{i=1}^{m}\lambda_i  y_i w.x_i   +   \displaystyle\sum_{i=1}^{m}\lambda_i \\
 
  & =  & \frac{1}{2} \displaystyle\sum_{i=1}^{m}\lambda_i y_i x_i. \displaystyle\sum_{j=1}^{m}\lambda_j y_j x_j - \displaystyle\sum_{i=1}^{m}\lambda_i  y_ix_i  .\displaystyle\sum_{j=1}^{m}\lambda_j y_j x_j   +   \displaystyle\sum_{i=1}^{m}\lambda_i \\
  
    & =  & - \frac{1}{2} \displaystyle\sum_{i=1}^{m}\lambda_i y_i x_i. \displaystyle\sum_{j=1}^{m}\lambda_j y_j x_j    +   \displaystyle\sum_{i=1}^{m}\lambda_i \\
    
        & =  &\displaystyle\sum_{i=1}^{m}\lambda_i  - \frac{1}{2} \displaystyle\sum_{i=1}^{m}\displaystyle\sum_{j=1}^{m}\lambda_i \lambda_j y_i y_j x_i. x_j     \\
\end{array}
\label{eqwolf4}
\end{equation}

The Wolfe dual problem becomes:

\begin{equation}
\begin{array}{ccc}
&\displaystyle \max_{\lambda} \displaystyle\sum_{i=1}^{m}\lambda_i  - \frac{1}{2} \displaystyle\sum_{i=1}^{m}\displaystyle\sum_{j=1}^{m}\lambda_i \lambda_j y_i y_j x_i. x_j &\\
&\textrm{s.t.} \lambda_i \geq 0 \wedge \displaystyle\sum_{i=1}^{m}\lambda_i  y_i = 0, i\in\{1,...,m\} &\\
\end{array}
\label{eqwolf5}
\end{equation}

Note that  the objective function in the Wolfe dual problem depends solely on the Lagrangian multipliers, which makes it easier to solve than the Lagrange dual problem.


\paragraph*{Soft Margin SVM} 

When the data are not  linearly separable due to noisy points, which is often the case, the above problem formulation could lead to nothing. For that, we introduce slack variables $\zeta_i$ and a regularization parameter $r$ such that optimization problem becomes

$$
\begin{array}{ccc}
&\displaystyle \min_{w,b,\xi}  {\frac{1}{2}||w||^2 + r.\displaystyle\sum_{j=1}^{m}\zeta_i} &\\
&\textrm{s.t.} y_i(w.x+b) -1 + \zeta_i\geq 0, \zeta_i \geq 0 \wedge i \in \{1 ... m\} &\\
\end{array}
$$

The variables $\zeta_i$ are introduced to tolerate points that do not satisfy the original constraints. The regularization parameter $r$ is introduced to control how much SVM will tolerate misclassification in each trained example. In fact, a high value of $r$ will reduce the impact of $\zeta_i$ (low misclassification tolerance), and a low value of $r$ will increase the impact of $\zeta_i$ (high misclassification tolerance).

\subsubsection*{Kernel tricks} 

Kernel tricks (or kernels) are introduced to allow an SVM to deal with data that are not linearly separable by nature and not due to noise (see Figure \ref{fig:SVM2}).

The trick is to find a mapping function $\varphi$ that transforms the dataset into a larger space where linear SVM can be performed (see Figure \ref{fig:SVM3}). 

Notice that in Wolfe dual problem given in \ref{eqwolf5}, the maximization problem hinges only on the dot product of the support vectors (i.e. $x_i.x_j$).

If we map the dataset  to a higher-dimensional space, we will have to maximize the objective function: $$ \displaystyle \max_{\lambda} \displaystyle\sum_{i=1}^{m}\lambda_i  - \frac{1}{2} \displaystyle\sum_{i=1}^{m}\displaystyle\sum_{j=1}^{m}\lambda_i \lambda_j y_i y_j\varphi( x_i). \varphi(x_j)$$
under the same constraints.

Finding the mapping function $\varphi$ is a hard task in practice. However, we do not really have to. All we need is the dot product $\varphi( x_i). \varphi(x_j)$ for the support vectors. Here the Mercer theorem \cite{mercer} comes in handy. It states that under some conditions on the form of a kernel function $K$ (symmetric, positive, and semi-definite), there exists a mapping function $\varphi$ such that we have always:

$$ K(x_i,x_j) = \varphi(x_i).\varphi(x_j)$$ 

The example at the end of this annex shows this is true for the Gaussian kernel function. The SVM maximization problem becomes

$$ \displaystyle \max_{\lambda} \displaystyle\sum_{i=1}^{m}\lambda_i  - \frac{1}{2} \displaystyle\sum_{i=1}^{m}\displaystyle\sum_{j=1}^{m}\lambda_i \lambda_j y_i y_j K(x_i,x_j)$$

As we can see, SVM tightly depends on the chosen kernel.

In Table \ref{tabKernels}, we give some common kernel functions \cite{kernel2}. 

\begin{table}[!htb]
\centering

\begin{tabular}{|L{5cm}|L{5cm}|}
\hline
\Z $K(x_i,x_j) = (x_i.x_j+1)^p$ &    Polynomial kernel \\\hline
\Z $K(x_i,x_j) = e^{-\displaystyle \frac{||x_i-x_j||^2}{2\sigma ^2}}$ &    Gaussian kernel \\\hline
\Z $K(x_i,x_j) = e^{-\displaystyle\gamma||x_i-x_j||^2}$ &    Gaussian Radial Basis Function (RBF) \\\hline
\Z $K(x_i,x_j) = e^{-\displaystyle\frac{||x_i-x_j||}{\sigma}}$ &    Laplace RBF kernel \\\hline
\Z $K(x_i,x_j) = tanh (\displaystyle\eta x_i.x_j +\nu)$ &    Hyperbolic tangent kernel \\\hline
\end{tabular}

\caption{Common SVM kernel functions}
\label{tabKernels}
\end{table}

\pagebreak

\textbf{Example:}
$ $\\

Let $K(x_i,x_j) = (x_i.x_j+1)^p$ be the Gaussian kernel function with $p=2$.

\begin{tabular}{l l L{12cm}}

\ZZ $K(x_i,x_j)$ & $=$ &  $K((x_{i_1},x_{i_2}), (x_{j_1},x_{j_2}))$ \\
\ZZ & $=$ &  $1+(x_{i_1}x_{j_1})^2 + (x_{i_2}x_{j_2})^2 + 2x_{i_1}x_{j_1}+2x_{i_2}x_{j_2} + 2x_{i_1}x_{j_1}x_{i_2}x_{j_2}$ \\
\ZZ & $=$ &  $(1   ,    \sqrt{2} x_{i_1}   ,      \sqrt{2} x_{i_2}    ,   x_i^2   ,   \sqrt{2} x_{i_1} x_{i_2}  , x_{i_2}^2)$ $.$ $(1   ,    \sqrt{2} x_{j_1}   ,      \sqrt{2} x_{j_2}    ,   x_i^2   ,   \sqrt{2} x_{j_1} x_{j_2}  , x_{j_2}^2)$\\
\ZZ & $=$ &  $\varphi(x_i).\varphi(x_j)$, where\\
\ZZ $\varphi(x_i)$& $=$ & $(1   ,    \sqrt{2} x_{i_1}   ,      \sqrt{2} x_{i_2}    ,   x_i^2   ,   \sqrt{2} x_{i_1} x_{i_2}  , x_{i_2}^2)$, and \\
\ZZ $\varphi(x_j)$ & $=$ &$(1   ,    \sqrt{2} x_{j_1}   ,      \sqrt{2} x_{j_2}    ,   x_i^2   ,   \sqrt{2} x_{j_1} x_{j_2}  , x_{j_2}^2)$
 \\

\end{tabular}




\chapter{ Common activation functions}

 
In artificial neural networks, a neuron does not directly output its weighted sum $b + \displaystyle \Sigma_{i}^{} w_i*x_i$. Instead, this sum passes through a function called an activation function (or  transfer function) that has a flattening effect on the output and transforms it to values like yes or no. It maps the resulting values to a span ranging from  $0$ to $1$ or $-1$ to $1$, etc. A neural network without an activation function is simply a linear regression model. The activation functions also make back-propagation possible since gradients are provided along with error to update weights and biases. 

There are two categories of activation functions: linear and non-linear. Linear activation functions are functions that are represented by a line. Their derivatives are constants. They are rapid and suitable when the addressed problem is linear by nature. Nonlinear activation functions are the most used activation functions. They are suitable for problems that are not linear by nature, which are most often encountered in Machine Learning. Table \ref{ActFUNC} presents the most utilized activation functions, namely sigmoid, hyperbolic tangent (tanh), rectified linear unit (ReLU), parametric rectified Linear unit (PReLU or Leaky ReLU), exponential linear unit (ELU), and  Softplus. Table \ref{ActFUNCProsCons} presents the advantages and drawbacks of these functions. 
It is important to also mention the Softmax function used in multi-class classification problems, which exponentiates the outputs and renormalizes them. This makes their sum equal to 1 and allows each output to be interpreted as a probability:

 $$ \sigma(x_i) = \displaystyle \frac{e^{x_i}}{\displaystyle \sum_{k=1}^{K}e^{x_k}}, \text{ for } i=1,..., K.$$

This mainly gives the probability of the output belonging to a particular class among the $K$ classes.\\ Further details on activation functions can be found in \cite{Nwankpa, FARHADI}.\\

\begin{table}[]

\centering 
\begin{tabular}{L{6cm} L{6cm}}

 \fbox{\includegraphics[scale=0.3]{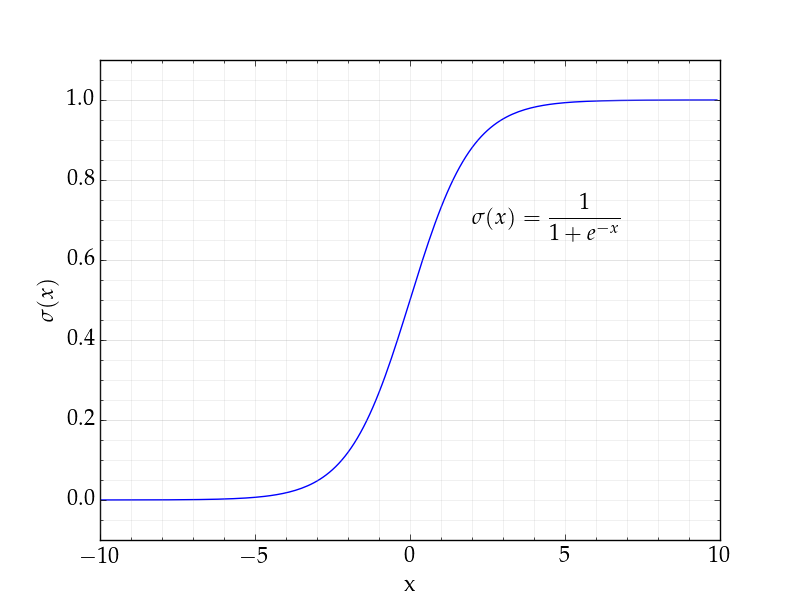}}
 & 
 \fbox{\includegraphics[scale=0.3]{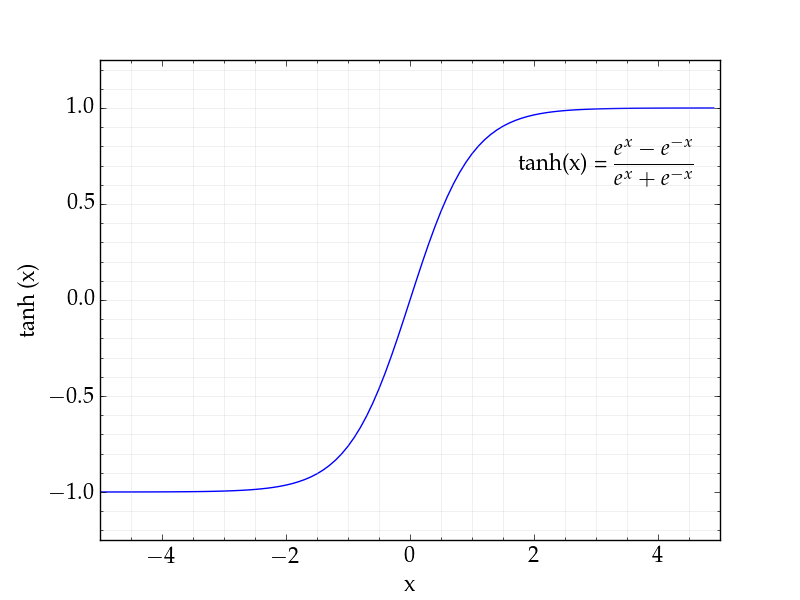}}\\

 \fbox{\includegraphics[scale=0.3]{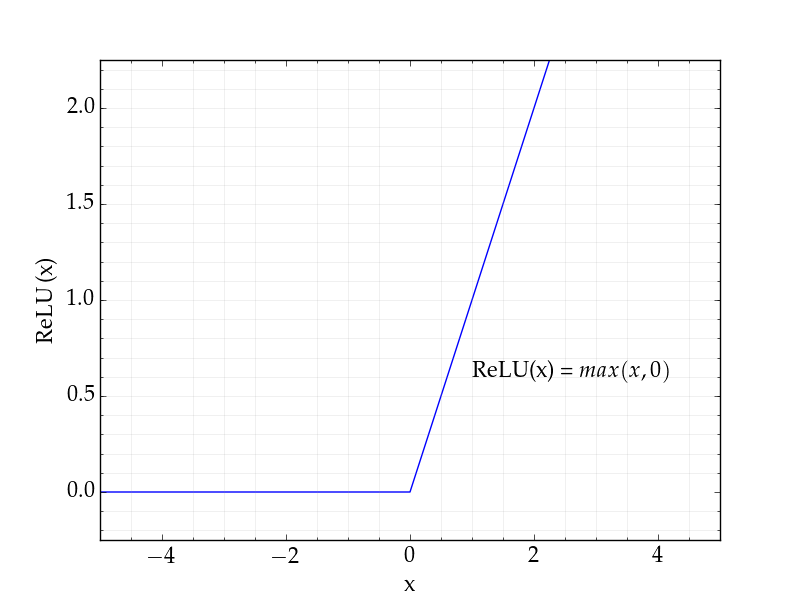}}
 & 
  \fbox{\includegraphics[scale=0.3]{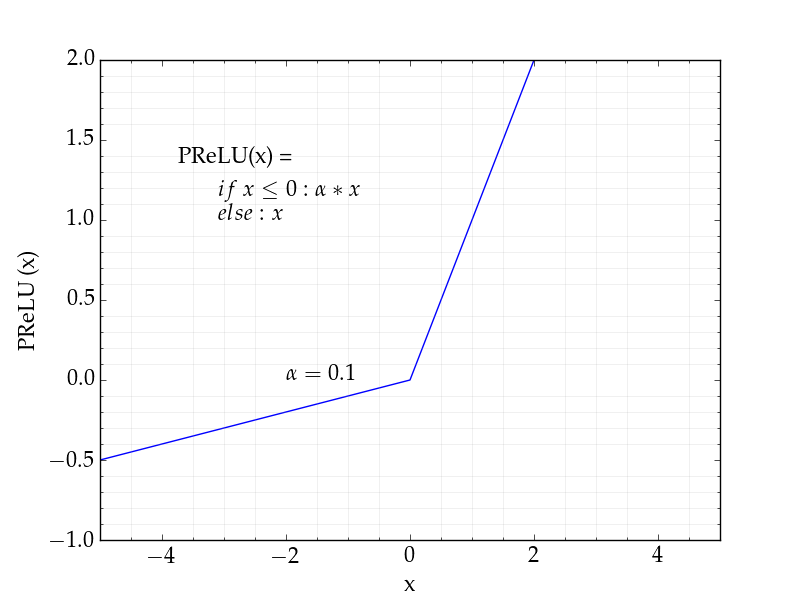}}\\

 \fbox{\includegraphics[scale=0.3]{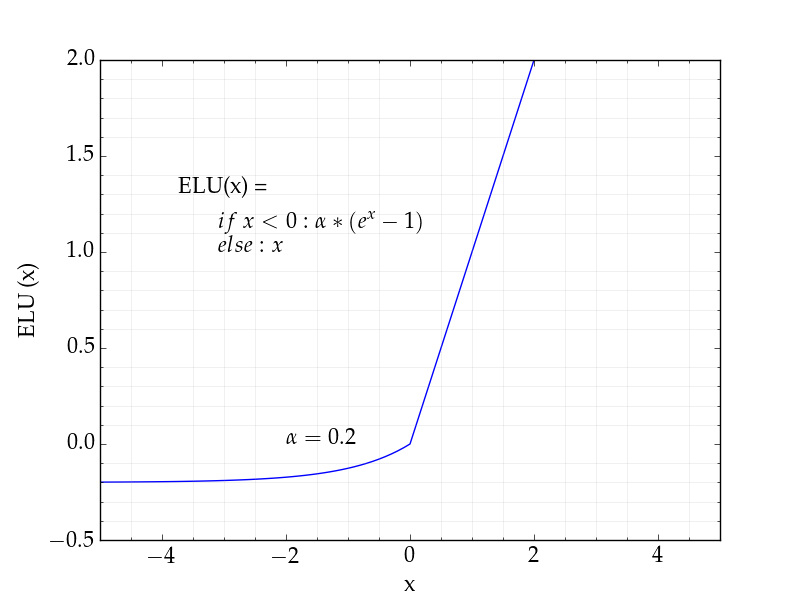}}
 & 
  \fbox{\includegraphics[scale=0.3]{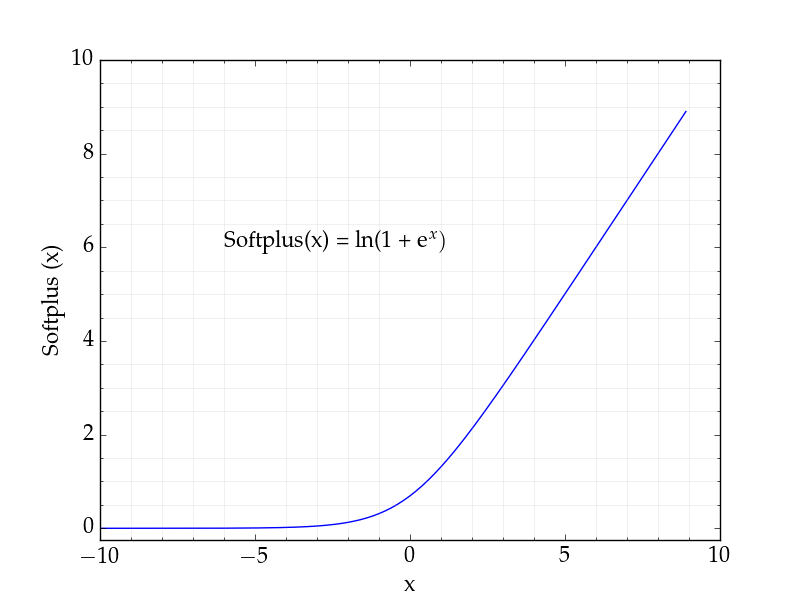}}\\

\end{tabular}
\caption{Common activation functions}
\label{ActFUNC}
\end{table}

\begin{table}
\begin{tabular}{|C{4cm}|C{4cm}|C{4cm}|}
\hline
\ZZ \textbf{Activation function} & \textbf{Advantages}& \textbf{Disadvantages}\\
\hline

\ZZ Sigmoid ($\sigma$) 
&    
differentiable, monotonic. 
& 
vanishing gradients, output is not zero-centered, saturation, slow convergence.
\\\hline

\ZZ Hyperbolic tangent (tanh)
&    
differentiable and monotonic, derivative is monotonic, output is zero-centered, easy to optimize.

& 
vanishing gradients, saturation, slow convergence.
\\\hline

\ZZ Rectified Linear Unit (ReLU)
&    
differentiable and monotonic,  derivative is monotonic, does not activate all the neurons at the same time, efficient.

& 
outputs are not zero-centered, when the gradient attains zero for the negative values, it does not converge towards the minimum.
\\\hline

\ZZ Parametric Rectified Linear Unit (PReLU or Leaky ReLU)
&    
differentiable and monotonic,  derivative is monotonic, efficient, allows negative value during back propagation.

& 
results could be inconsistent, outputs are not zero-centered, if the learning rate is very high this will overshoot killing the neuron.
\\\hline

\ZZ Exponential Linear Unit (ELU) 
&    
improves training performance, uses exponential for the negative part and no longer a linear function, performs better than the other ReLUs, no worries about dead neuron (dying ReLU).

& 
for positive values, it may blow up the activation (it ranges between $[0, \infty[$)
\\\hline

\ZZ Softplus
&    
differentiable and monotonic,  the derivative of  softplus  is the sigmoid,  efficient, used for multivariate classification.

& 
vanishing gradients, output is not zero centered, saturation
\\\hline

\end{tabular}

\caption{Activation functions: advantages and disadvantages}
\label{ActFUNCProsCons}
\end{table}

\chapter{Gradient descent optimizers}

The role of the gradient descent is not only to find the minimum of a cost function (also called loss function or objective function), but also to optimize the convergence to that minimum. Gradient descent uses optimizers for that purpose. In fact, the descent to the optimum may be very expensive in terms of speed of convergence (slow). For that, optimizers could be very useful in accelerating that convergence, but they may negatively affect the model accuracy. A trade-off between speed and accuracy should be negotiated.

\subsection*{Notations}

In this annex, we use the following notations, which are the most common notations used in machine learning literature.


\begin{itemize}
\item $\theta$ is a vector of $\displaystyle\mathbb{R}^d$. It represents the model parameters (weights and biases vector).\\

\item $J(\theta)$:  is the cost function. 

\item The gradient descent updates the parameters $\theta$ in the direction opposite that of the gradient of the cost function, denoted by $\nabla J(\theta)$. \\

\item $\eta$ is the learning rate which represents the size of the steps the descent takes towards the minimum.\\

\item $x^{(i)}$ and $y^{(i)}$ are respectively the $i$th training example and its label. When we do not need to specify the rank $i$, it is omitted.

\end{itemize}

\section*{Gradient descent types}

There are three types of gradient descents which differ in the amount of data used to calculate the gradient of the cost function.

\begin{itemize}

\item     Batch Gradient Descent or Vanilla Gradient Descent: uses the entire dataset. The gradient is calculated as follows

$$\theta = \theta - \eta \nabla J(\theta)$$

and the descent makes one update on the model parameters every epoch.

\item     Stochastic Gradient Descent (SGD):  makes one parameter update at a time, for every single training example $x^{(i)}$ and $y^{(i)}$.  The gradient is calculated as follows

$$\theta = \theta - \eta \nabla J(\theta, x^{(i)}, y^{(i)})$$

\item     Mini-batch Gradient Descent: makes one parameter update for every $n$ training examples $x^{(i:i+n)}$ and $y^{(i:i+n)}$.  The gradient is calculated as follows:

$$\theta = \theta - \eta \nabla J(\theta, x^{(i:i+n)}, y^{(i:i+n)})$$

\end{itemize}

\pagebreak

\begin{table}
\begin{tabular}{|C{4cm}|C{4cm}|C{4cm}|}
\hline
\ZZ \textbf{Name} &  \textbf{Advantages}  & \textbf{Disadvantages}\\
\hline
\ZZ Batch Gradient Descent 

&   

One update at a time

& 
Slow gradient descent; 
difficult for datasets that don't fit in memory;
impossible for online learning; 
high variance

\\ \hline

\ZZ Stochastic Gradient Descent (SGD) &  
Sure convergence for both non-convex and convex optimization; 
suitable for datasets that don't fit in memory;
possible online learning
  & 
High variance;
fluctuates heavily

\\ \hline

\ZZ Mini-batch Gradient Descent 
&    
Reduces the variance; 
stable convergence
& 
Programmer should find the best batch size
\\ \hline

\end{tabular}

\caption{Comparison between gradient descent types}
\label{tabGradientTypes}
\end{table}

\subsection*{Gradient descent optimization algorithms}

\subsubsection*{Momentum}

Momentum \cite{QIAN1999145} is an algorithm to accelerate SGD. It \textit{pushes} the descent down the hill in the relevant direction and mitigates oscillations in irrelevant directions. Momentum adds a fraction $\gamma$ of the previous update vector to the current update vector as given by:

$$
 \left\{
    \begin{array}{ll}
        v_t = \gamma v_{t-1} + \eta \nabla J(\theta)& \\
\\
        \theta = \theta - v_{t-1}& \\
        
    \end{array}
\right.
$$

$\gamma$ is usually set close to $0.9$.

\subsection*{Nesterov accelerated gradient}

The Nesterov accelerated gradient (NAG) algorithm \cite{YuriiNesterov} is similar to Momentum with the difference that the corrective term $v_t$ approximates the next position of the parameters (by computing $\theta - \gamma v_{t-1}$) prior to performing the current update. This is given by:

$$
 \left\{
    \begin{array}{ll}
        v_t = \gamma v_{t-1} + \eta \nabla J(\theta - \gamma v_{t-1})& \\
\\
        \theta = \theta - v_{t}& \\
        
    \end{array}
\right.
$$

This algorithm prevents blind descent (uncontrolled descent that may diverge) and gives the Momentum algorithm more orientation toward the minimum.

\subsection*{AdaGrad}


AdaGrad \cite{Duchi:2011:ASM:1953048.2021068} is an adaptive algorithm that makes small updates to parameters related to frequently occurring features and larger updates to parameters related to less frequent features by scaling the learning rate for each dimension.  The equation for the parameter update  is as follows:

$$\theta_{t+1}^{(i)} = \theta_{t}^{(i)} - \displaystyle \frac{\eta}{\sqrt{\epsilon + G_{t}^{(i,i)}}}g_{t}^{(i)}$$

where $\epsilon$ is a small term added to prevent the division by zero, $g_{t,i}$ is the gradient of the loss function at the time step $t$ for the dimension $i$, and $G_{t}^{(i,i)}$ is the sum of each step's squared gradients up until the current time step (i.e. $G_{t}^{(i,i)} = \displaystyle \sum_{\tau = 1}^{t} (g_{\tau}^{(i)})^2$).

The main drawback of AdaGrad is the accumulation of the squared gradients in the denominator (positive numbers), which may cause the algorithm to stop learning.

\subsubsection*{Adadelta}

The Adadelta  algorithm \cite{DBLP:journals/corr/abs-1212-5701} is devised to overcome the insufficiencies of  AdaGrad by limiting the accumulated past gradients to a time window of a fixed size. Its parameter update equation is as follows:

$$\theta_{t+1}^{(i)} = \theta_{t}^{(i)} - \displaystyle \frac{RMS[\Delta \theta]_{t-1}^{(i)}}{RMS[g]_{t}^{(i)}}g_{t}^{(i)}$$

where  $RMS[\Delta \theta]_{t-1}^{(i)}$ is the root mean squared error of the parameter updates until the previous time step in the time window and $RMS[g]_{t}^{(i)}$  is  the root mean squared error of the gradients until the current time step  in the time window.

\subsubsection*{Adam}

Adaptive Moment estimation (Adam) \cite{DBLP:journals/corr/KingmaB14} calculates two quantities $m_t$ and $v_t$ before updating the parameters as follows:

$$
 \left\{
    \begin{array}{ll}
        m_t = \beta_1 m_{t-1} + (1-\beta_1) g_t& \\
\\
        v_t  = \beta_2v_{t-1} + (1- \beta_2)|g_t|^2& \\
        
    \end{array}
\right.
$$

where $g_t$ is the gradient of the loss function, and $\beta_1$ and $\beta_2$ are two factors close to $1$. $m_t$ is an estimate of the first moment and $v_t$ is an estimate of the second moment representing the uncentered variance.

It turns out that these two quantities are biased towards zero. To remedy to that, the estimates of the first moment and the second moment are respectively approximated as follows:

$$
 \left\{
    \begin{array}{ll}
        \hat{m}_t = \displaystyle \frac{m_t}{1-\beta_1^t}& \\
\\
       \hat{v}_t = \displaystyle \frac{v_t}{1 - \beta_2^t} & \\
        
    \end{array}
\right.
$$

Finally,  the parameters are updated as follows:

$$\theta_{t+1} = \theta_t - \displaystyle \frac{\eta}{\displaystyle \sqrt{\hat{v}} + \epsilon} \hat{m}_t$$

where $\epsilon$ is a positive number close to zero to prevent division by zero.

\subsubsection*{AdaMax}

Unlike Adam, AdaMax \cite{DBLP:journals/corr/KingmaB14} uses the infinite norm $l_{\infty}$ rather than the norm $l_2$ when calculating $v_t$ and replaces the term ${\displaystyle \sqrt{\hat{v}} + \epsilon}$ with $u_t = \texttt{max} (\beta_2 v_{t-1}, |g_t|)$. The convergence turns out to be more stable in many cases when using the infinite norm.

The parameters are updated as follows:

$$\theta_{t+1} = \theta_t - \displaystyle \frac{\eta}{u_t} \hat{m}_t$$

\subsubsection*{Nadam}

Nesterov-accelerated Adaptive Moment estimation (Nadam) \cite{DBLP:journals/corr/Ruder16} is a combination of Adam and NAG with some approximations. It updates the parameters as follows:

$$\theta_{t+1} = \theta_t - \displaystyle \frac{\eta}{\displaystyle \sqrt{\hat{v}} + \epsilon} (\beta_1 \hat{m}_t + \displaystyle \frac{1 - \beta_1}{1 - \beta_1^t} g_t)$$

where all the parameters are the same as in Adam.

\subsubsection*{AMSGrad}

AMSGrad \cite{amsgrad, DBLP:journals/corr/abs-1904-03590} is to some extent a combination of Adam, AdaMax, and AdaGrad. It uses the maximum of past squared gradients to update the parameters as follows:

$$
 \left\{
    \begin{array}{ll}
       m_t = \beta_1 m_{t-1} + (1-\beta_1) g_t & \\
\\
       v_t = \beta_2 v{t-1} + (1 - \beta_2) |g_t|^2& \\
\\
      \hat{v} = \texttt{max}({\hat{v}_{t-1}, v_t)}& \\
\\

       \theta_{t+1} = \theta_t - \displaystyle \frac{\eta}{\displaystyle \sqrt{\hat{v}} + \epsilon}  {m}_t & \\
        
    \end{array}
\right.
$$

where all the parameters are the same as in Adam, AdaMax, and AdaGrad.

\subsubsection*{RMSprop}

RMSprop \cite{RMSprop} is very similar to AdaGrad. It also uses the past squared gradients to update the parameters as follows:

$$
 \left\{
    \begin{array}{ll}
       E[g_t^2] = \beta  E[g^2]_{t-1} + (1-\beta) g_t^2& \\
\\
      \theta_{t+1} = \theta_t - \displaystyle \frac{\eta}{\displaystyle \sqrt{E[g_t^2]} + \epsilon}  {m}_t& \\       
    \end{array}
\right.
$$

where all the parameters are the same as in AdaGrad and $E(g_t^2)$ is the moving average of squared gradients at time step $t$.

\subsection*{Which is the best optimizer?}

It is difficult to answer this question! 

This tightly depends on the data in the training dataset. The state of the art at the present time lacks reliable results regarding the preference of one optimizer over another. Hence, the choice of the most adequate optimizer for a given problem remains an empirical observation.

\end{document}